\documentclass[12pt]{article}

\usepackage{graphicx}
\usepackage{amssymb}
\usepackage{accents}

\newtheorem{theorem}{Theorem}
\newtheorem{coro}{Corollary}

\newcommand{\bfa}{\mathbf{a}} 

\newcommand{\bfc}{\mathbf{c}}
\newcommand{\bfd}{\mathbf{d}}  
\newcommand{\bfe}{\mathbf{e}}  
\newcommand{\bff}{\mathbf{f}}  

\newcommand{\bfm}{\mathbf{m}}
\newcommand{\bfu}{\mathbf{u}}
\newcommand{\bfx}{\mathbf{x}}
\newcommand{\bfy}{\mathbf{y}}
\newcommand{\bfz}{\mathbf{z}}

\newcommand{\bfA}{\mathbf{A}} 
\newcommand{\bfB}{\mathbf{B}}
\newcommand{\bfC}{\mathbf{C}}

\newcommand{\bfF}{\mathbf{F}}

\newcommand{\bfI}{\mathbf{I}}
\newcommand{\bfJ}{\mathbf{J}}
\newcommand{\bfK}{\mathbf{K}}
\newcommand{\bfL}{\mathbf{L}}
\newcommand{\bfM}{\mathbf{M}}
\newcommand{\bfR}{\mathbf{R}}
\newcommand{\bfT}{\mathbf{T}}
\newcommand{\bfU}{\mathbf{U}}
\newcommand{\bfV}{\mathbf{V}}
\newcommand{\bfW}{\mathbf{W}}

\newcommand{\bfcg}{\accentset{\frown}{\mathbf{c}}}
\newcommand{\bfeg}{\accentset{\frown}{\mathbf{e}}}
\newcommand{\bffg}{\accentset{\frown}{\mathbf{f}}}
\newcommand{\bfxg}{\accentset{\frown}{\mathbf{x}}}
\newcommand{\bfGg}{\accentset{\frown}{\mathbf{G}}}

\newcommand{\bfTh}{\mathbf{\Theta}}
\newcommand{\zeros}{\mathbf{0}}
\newcommand{\sM}{\mathcal{M}}

\begin{document}




\title{Stable Concurrent Synchronization \\ in Dynamic System Networks}

\author{{\normalsize Quang-Cuong Pham} \\
{\normalsize D\'epartement d'Informatique} \\
{\normalsize \'Ecole Normale Sup\'erieure} \\
{\normalsize Paris, France} \\
{\normalsize \texttt{cuong.pham@ens.fr}}
\and
{\normalsize Jean-Jacques Slotine }\\
{\normalsize Nonlinear Systems Laboratory }\\
{\normalsize Massachusetts Institute of Technology }\\
{\normalsize Cambridge, MA 02139, USA }\\
{\normalsize \texttt{jjs@mit.edu}}
}

\maketitle





\begin{abstract}
  In a network of dynamical systems, concurrent synchronization is a
  regime where multiple groups of fully synchronized elements
  coexist. In the brain, concurrent synchronization may occur at
  several scales, with multiple ``rhythms'' interacting and functional
  assemblies combining neural oscillators of many different
  types. Mathematically, stable concurrent synchronization corresponds
  to convergence to a flow-invariant linear subspace of the global
  state space. We derive a general condition for such convergence to
  occur globally and exponentially. We also show that, under mild
  conditions, global convergence to a concurrently synchronized regime
  is preserved under basic system combinations such as negative
  feedback or hierarchies, so that stable concurrently synchronized
  aggregates of arbitrary size can be constructed. Robustnesss of
  stable concurrent synchronization to variations in individual
  dynamics is also quantified. Simple applications of these results to
  classical questions in systems neuroscience and robotics are
  discussed.
\end{abstract}



\section{Introduction}

Distributed synchronization phenomena are the subject of intense
research. In the brain, such phenomena are known to occur at different
scales, and are heavily studied at both the anatomical and
computational levels.  In particular, synchronization has been
proposed as a general principle for temporal binding of multisensory
data~\cite{SinGra,Grossberg,Llinas,Mount,Tononi,Koch,NieEbi}, and as a
mechanism for perceptual grouping~\cite{YazGro}, neural
computation~\cite{Braitenberg,Brody,WangSlo5} and neural
communication~\cite{Kandel,IzhiDes,Schnitzler,Schoffelen}. Similar
mathematical models describe fish schooling or certain types of
phase-transition in physics~\cite{Strogatz}.

In an ensemble of dynamical elements, {\it concurrent synchronization}
is defined as a regime where the whole system is divided into multiple
groups of fully synchronized elements\footnote{In the literature, this
phenomenon is often called \emph{poly-}, or \emph{cluster} or
\emph{partial} synchronization. However, the last term can also
designate a regime where the elements are not fully synchronized but
behave coherently~\cite{Strogatz}.}, but elements from different
groups are not necessarily synchronized~\cite{Belykh1,Zhang,PogSanNij}
and can be of entirely different dynamics~\cite{GolSte}. It can be
easily shown that such a regime corresponds to a flow-invariant linear
subspace of the global state space. Concurrent synchronization
phenomena are likely pervasive in the brain, where multiple
``rhythms'' are known to coexist~\cite{Kandel,Schnitzler}, neurons can
exhibit many qualitatively different types of
oscillations~\cite{Kandel,Izhi}, and functional models often combine
multiple oscillatory dynamics.

In this paper, we introduce a simple sufficient condition for a
general dynamical system to converge to a flow-invariant subspace. Our
analysis is built upon nonlinear contraction
theory~\cite{LohSlo,WangSlo}, and thus it inherits many of the
theory's features :

\begin{itemize}
\item global exponential convergence and stability are guaranteed,
\item convergence rates can be explicitly computed as
  eigenvalues of well-defined symmetric matrices,
\item robustness to variations in dynamics can be easily quantified,
\item under simple conditions, convergence to a concurrently 
  synchronized state can be preserved through system combinations.
\end{itemize}


As we shall see, under simple conditions on the coupling strengths,
architectural symmetries~\cite{GolSteTor} and/or diffusion-like
couplings \emph{create globally stable concurrent synchronization
  phenomena}. This is illustrated in figure 1, which is very loosely
inspired by oscillations in the thalamocortical
system~\cite{Llinas,SinGra,Koch,NieEbi,Schnitzler}. Qualitatively,
global stability of the concurrent synchronization is in the same
sense that an equilibrium point is globally stable -- any initial
conditions will lead back to it, in an exponential fashion.  But of
course it can yield extremely complex, coordinated behaviors.

\begin{figure}[ht]
  \centering
  \includegraphics[scale=0.6]{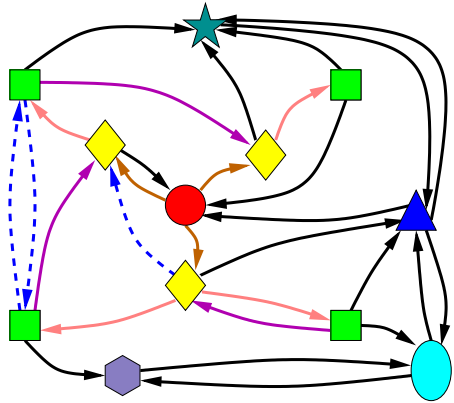}
  \caption[]{\small An example of concurrent synchronization.
    Systems and connections of the same shape (and color) have
    identical dynamics (except black arrows, which represent arbitrary
    connections, and dashed arrows, which represent diffusive
    connections). This paper shows that under simple conditions on the
    coupling strengths, the group of (green) squares globally exponentially 
    synchronizes (thus providing synchronized input to the outer elements), 
    and so does the group of (yellow) diamonds, regardless of 
    the specific dynamics, connections, or inputs of the other systems.} 
  \label{fig:thalamus}
\end{figure}

Section 2 recalls key concepts of nonlinear contraction theory and
derives a theoretical tool for studying global convergence to a
flow-invariant subspace. Section 3 presents the paper's main
mathematical results, relating stable concurrent synchronization to
coupling structures and flow-invariant subspaces created by symmetries
or diffusion-like couplings.  Robustnesss of concurrent
synchronization to variations in individual dynamics is also
quantified, showing in particular how approximate symmetries lead to
quasi-synchronization.  Section 4, motivated by evolution and
development, studies conditions under which concurrent synchronization
can be preserved through combinations of multiple concurrently
synchronized regimes.
Finally, section 5 discusses potential applications of these results
to general questions in systems neuroscience and robotics.



\section{Basic Tools}

\subsection{Nonlinear contraction theory} 
\label{sec:contraction}

This section reviews basic results of nonlinear contraction
theory~\cite{LohSlo,LohSlo2,Slo,WangSlo}, which is the main stability
analysis tool used in the paper. Essentially, a nonlinear time-varying
dynamic system will be called {\it contracting} if initial conditions
or temporary disturbances are forgotten exponentially fast, i.e., if
trajectories of the perturbed system return to their nominal behavior
with an exponential convergence rate.  It turns out that relatively
simple algebraic conditions can be given for this stability-like
property to be verified, and that this property is preserved through
basic system combinations.

While we shall derive global properties of nonlinear systems, many of
our results can be expressed in terms of eigenvalues of symmetric
matrices~\cite{horn}. Given a square matrix $\bfA$, the symmetric part
of $\bfA$ is denoted by $\bfA_s$. The smallest and largest eigenvalues
of $\bfA_s$ are denoted by $\lambda_\mathrm{min}(\bfA)$ and
$\lambda_\mathrm{max}(\bfA)$. Given these notations, the matrix $\bfA$
is {\it positive definite} (denoted $\bfA > \zeros$) if
$\lambda_\mathrm{min}(\bfA)>0$, and it is {\it negative definite}
(denoted $\bfA < \zeros$) if $\lambda_\mathrm{max}(\bfA)<0$.  Finally,
a square matrix $\bfA(\bfx,t)$ is \emph{uniformly} positive definite
if \mbox{$\exists \beta >0, \forall \bfx, \forall t :
\lambda_\mathrm{min}(\bfA(\bfx,t)) \ge \beta $}, and it is \emph{uniformly}
negative definite if \mbox{$\exists \beta >0, \forall \bfx, \
\forall t : \lambda_\mathrm{min}(\bfA(\bfx,t)) \le -\beta$}.

The basic theorem of contraction analysis, derived in~\cite{LohSlo},
can be stated as:
\begin{theorem}[Contraction]
\label{theorem:contraction}
Consider, in $\mathbb{R}^n$, the deterministic system
\begin{equation}
  \label{equ:main2}
  \dot\bfx = \bff(\bfx,t)
\end{equation}
where $\bff$ is a smooth nonlinear function. Denote the Jacobian
matrix of $\bff$ with respect to its first variable by $\frac{\partial
  \bff} {\partial\bfx}$. If there exists a square matrix
$\bfTh(\bfx,t)$ such that $\bfTh(\bfx,t)^\top\bfTh(\bfx,t)$ is
uniformly positive definite and the matrix
\[
\bfF = \left(\dot\bfTh + \bfTh \frac{\partial \bff} {\partial\bfx}
\right) \bfTh^{-1} 
\]
is uniformly negative definite, then all system trajectories converge
exponentially to a single trajectory, with convergence rate
$|\sup_{\bfx,t}\lambda_\mathrm{max}(\bfF)|>0$. The system is said to
be \emph{contracting}, $\bfF$ is called its \emph{generalized
Jacobian}, and $\bfTh(\bfx,t)^\top\bfTh(\bfx,t)$ its contraction
\emph{metric}.
\end{theorem}

It can be shown conversely that the existence of a uniformly positive
definite metric $ \bfM(\bfx,t)=\bfTh(\bfx,t)^\top\bfTh(\bfx,t)$ with
respect to which the system is contracting is also a necessary
condition for global exponential convergence of
trajectories~\cite{LohSlo}.  Furthermore, all transformations $\bfTh$
corresponding to the same $\bfM$ lead to the same eigenvalues for the
symmetric part $\bfF_s $ of $\bfF$~\cite{Slo}, and thus to the same contraction
rate $|\sup_{\bfx,t}\lambda_\mathrm{max}(\bfF)|$.

In the linear time-invariant case, a system is globally contracting if
and only if it is strictly stable, and $\bfF$ can be chosen as a
normal Jordan form of the system with $\bfTh$ the coordinate
transformation to that form~\cite{LohSlo}.  Contraction analysis can
also be derived for discrete-time systems and for classes of hybrid
systems~\cite{LohSlo2}.

Finally, it can be shown that contraction is preserved through basic system
combinations (such as parallel combinations, hierarchies, and certain
types of negative feedback, see~\cite{LohSlo} for details), a property
which we shall extend to the synchronization context in this paper
(section 4).

\begin{theorem}[Contraction and robustness]
  \label{theorem:robust}
  Consider a contracting system $\dot\bfx = \bff(\bfx,t)$, with $\bfTh
  = \bfI$ and contraction rate $\lambda$. Let $P_1(t)$ be a
  trajectory of the system, and let $P_2(t)$ be a trajectory of the
  \emph{disturbed} system
  \[
  \dot\bfx = \bff(\bfx,t) + \bfd(\bfx,t)
  \]
  Then the distance $R(t)$ between $P_1(t)$ and $P_2(t)$ verifies
$ R(t)\leq \sup_{\bfx,t}\|\bfd(\bfx,t)\| / \lambda$
 after exponential transients of rate $\lambda$.
\end{theorem}

For a proof and generalisation of this theorem, see section 3.7 in
\cite{LohSlo}.

\subsection{Convergence to a flow-invariant subspace}
\label{sec:convergence}

We now derive a simple tool upon which the analyses of this paper will
be based. The derivation is inspired by the idea of ``partial''
contraction, introduced in~\cite{WangSlo}, which consists in
applying contraction tools to virtual auxiliary systems so as to
address questions more general than trajectory convergence.

Consider again, in $\mathbb{R}^n$, the deterministic system
\begin{equation} 
  \label{equ:main}
  \dot{\bfx} = \bff(\bfx,t)
\end{equation}
where $\bff$ is a smooth nonlinear function. Assume that there exists
a {\it flow-invariant linear subspace} $\sM$ (i.e. a linear subspace
$\sM$ such that $\forall t : \bff(\sM,t)\subset \sM$), which implies
that any trajectory starting in $\sM$ remains in $\sM$. Let
$p=\textrm{dim}(\sM)$, and consider an orthonormal basis
$(\bfe_1,\dots,\bfe_n)$ where the first $p$ vectors form a basis of
$\sM$ and the last $n-p$ a basis of $\sM^\perp$. Define an
$(n-p)\times n$ matrix $\bfV$ whose rows are
$\bfe_{p+1}^\top,\dots,\bfe_n^\top$. $\bfV$ may be regarded as a
projection\,\footnote{For simplicity we shall call $\bfV$ a
  ``projection'', although the actual projection matrix is in fact
  $\bfV^\top\bfV$.} on $\sM^\perp$, and it
verifies~\cite{horn,JadMotBar} : \[\bfV^\top\bfV + \bfU^\top\bfU =
\bfI_n \quad \quad \quad \bfV\bfV^\top = \bfI_{n-p}\quad \quad \quad
\quad \bfx\in\sM \iff \bfV\bfx=\zeros
\]
where $\bfU$ is the matrix formed by the first $p$ vectors.

\begin{figure}[ht]
  \centering
  \includegraphics[scale=0.6]{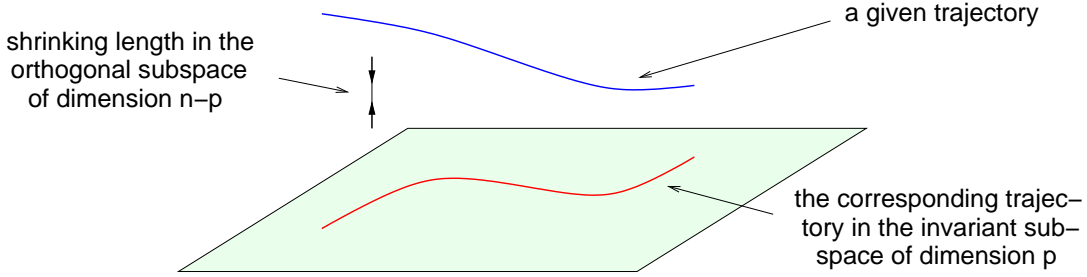}
  \caption{Convergence to a linear flow-invariant subspace}
  \label{fig:convergence}
\end{figure}

\begin{theorem}
  \label{theorem:convergence}
  Consider a linear flow-invariant subspace $\sM$ and the associated
  orthonormal projection matrix $\bfV$. A particular solution
  $\bfx_p(t)$ of system~(\ref{equ:main}) converges exponentially to
  $\sM$ if the system
  \begin{equation}
    \label{equ:sysy}
    \dot{\bfy}=\bfV\bff(\bfV^\top\bfy+\bfU^\top\bfU\bfx_p(t),t)
  \end{equation} is contracting with respect to $\bfy$. 

  If the above contraction condition is fullfilled for all $\bfx_p$,
  then starting from any initial conditions, all trajectories of
  system (\ref{equ:main}) will exponentially converge to $\sM$. If
  furthermore all the contraction rates for (\ref{equ:sysy}) are
  lower-bounded by some $\lambda>0$, uniformly in $\bfx_p$ and in a
  common metric, then the convergence to $\sM$ will be exponential
  with rate $\lambda$ (see figure \ref{fig:convergence}).

\end{theorem}

\textbf{Proof :} Let $\bfz_p=\bfV\bfx_p$. By construction, $\bfx_p$
converges to the subspace $\sM$ if and only if $\bfz_p$ converges to
$\zeros$.  Multiplying (\ref{equ:main}) by $\bfV$ on the left, we get
\begin{equation}
  \label{equ:sysz}
  \dot{\bfz}_p=\bfV\bff(\bfV^\top\bfz_p+\bfU^\top\bfU\bfx_p,t) 
\end{equation}


From (\ref{equ:sysz}), $\bfy(t)=\bfz_p(t)$ is a particular solution of
system (\ref{equ:sysy}). In addition, since
$\bfU^\top\bfU\bfx_p\in\sM$ and the linear subspace $\sM$ is
flow-invariant, one has
$\bff(\bfU^\top\bfU\bfx_p)\in\sM=\textrm{Null}(\bfV)$, and hence
$\bfy(t)=\zeros$ is another particular solution of system
(\ref{equ:sysy}). If system (\ref{equ:sysy}) is contracting with
respect to $\bfy$, then all its solutions converge exponentially to a
single trajectory, which implies in particular that $\bfz_p(t)$
converges exponentially to $\zeros$.

The remainder of the theorem is immediate.\ $\Box$


\begin{coro} 
  A simple sufficient condition for global
  exponential convergence to $\sM$ is that
  \begin{equation}
    \bfV\frac{\partial\bff}{\partial\bfx}\bfV^\top
    < \zeros \quad \rm{uniformly}
  \end{equation}
  or more generally, that there exists a constant invertible transform
  $\bfTh$ on $\sM^\perp$ such that
  \begin{equation}\label{metric}
    \bfTh\bfV\frac{\partial\bff}{\partial\bfx}\bfV^\top\bfTh^{-1}
    < \zeros \quad \rm{uniformly}
  \end{equation}
\end{coro}

\textbf{Proof :} The Jacobian of (\ref{equ:sysy}) with respect to
$\bfy$ is
\[ 
\bfV \left[ \frac{\partial\bff}{\partial\bfx}
  \left(\bfV^\top\bfy+\bfU^\top\bfU\bfx_p(t),t\right) \right]\bfV^\top
\]
so that the result is immediate by applying Theorem~\ref{theorem:contraction}.\ $\Box$


{\bf Remarks}
\begin{itemize}

\item \emph{Non-orthonormal bases.}  In practice, the subspace $\sM$
  is often defined by the conjunction of $(n-p)$ linear constraints.
  In a synchronization context, for instance, each of the constraints
  may be, e.g., of the form $\bfx_i = \bfx_j$ where $\bfx_i$ and
  $\bfx_j$ are subvectors of the state $\bfx$.  This provides directly
  a (generally not orthonormal) basis ($\bfe'_{p+1},\dots,\bfe'_n $)
  of $\sM^\perp$, and thus a matrix $\bfV'$ whose rows are
  ${\bfe'_{p+1}}^\top,\dots,{\bfe'_n}^\top$, and which verifies
  \[\bfV' = \bfT \bfV
  \]
  with $\bfT$ an invertible $(n-p)\times (n-p)$ matrix. We have $\
  \bfx\in\sM \iff \bfV'\bfx=\zeros$ and 
  \begin{equation}
    \label{equ:v'}
    \bfV\frac{\partial\bff}{\partial\bfx}\bfV^\top
    < \zeros \iff
    \bfV'\frac{\partial\bff}{\partial\bfx}{\bfV'}^\top
    < \zeros
  \end{equation}
  
  Consider for instance three systems of dimension $m$ and two systems
  of dimension $p$, and assume that $\sM =\{\bfx_1=\bfx_2,
  \bfx_5= - 10\bfx_4\}$ is the synchronization subspace of interest (with
  $\bfx_i $ denoting the state of each individual system).  One has
  directly
  \[
  \bfV'=\left(
    \begin{array}{ccccc}
      \bfI_m&-\bfI_m&\zeros&\cdots &\zeros\\
      \zeros&\cdots&\zeros&10\bfI_p& \bfI_p
    \end{array}\right)
  \]
  
  Note however that the equivalence in equation (\ref{equ:v'}) does
  not yield the same upper bound for the eigenvalues of the two
  matrices. Thus, in order to compute explicitly the convergence rate
  to $\sM$, one has to revert to the orthonormal version, using e.g. a
  Gram-Schmidt procedure~\cite{horn} on the rows of $\bfV'$.
  
\item \emph{More general invariant subspaces.}  This theorem can be
  extended straightforwardly to time-varying affine invariant
  subspaces of the form $\bfm(t)+\sM$ (apply the theorem to
  $\tilde{\bfx}(t)=\bfx(t)-\bfm(t)$). Preliminary results have also
  been obtained for nonlinear invariant manifolds
  \cite{PhamSloReport}.
\end{itemize}





\subsection{Global synchronization in networks of coupled identical
  dynamical elements}

\label{sec:global}

In this section, we provide by using theorem \ref{theorem:convergence}
a unifying and systematic view on several prior results in the study
of synchronization phenomena (see
e.g.~\cite{JadLinMo,PogSanNij,WangSlo,OlfMur,LinBroFra} and references
therein).

Consider first a network containing $n$ identical dynamical elements
with \emph{diffusive} couplings~\cite{WangSlo}
\begin{equation}
  \label{equ:sync}
  \dot{\bfx}_i = \bff(\bfx_i,t) + \sum_{j \neq i}
  \bfK_{ij}(\bfx_j - \bfx_i)    
  \qquad i=1,\ldots,n
\end{equation}

Let $\bfL$ be the Laplacian matrix of the network ($\bfL_{ii}=\sum_{j
  \neq i} \bfK_{ij}$, $\bfL_{ij}=-\bfK_{ij}$ for $j\neq i$),
and\,\footnote{The overscript $\frown$ denotes a vector in the global
  state space, obtained by grouping together the states of the
  elements.}
\[
\bfxg=\left(
\begin{array}{c}
\bfx_1\\
\vdots\\
\bfx_n
\end{array}\right), \quad
\bffg(\bfxg,t)=\left(
\begin{array}{c}
  \bff(\bfx_1,t)\\
  \vdots\\
  \bff(\bfx_n,t)
\end{array}\right)
\]

Equation (\ref{equ:sync}) can be rewritten in matrix form 
\begin{equation}
  \label{equ:syncx}
  \dot{\bfxg}=\bffg(\bfxg,t)-\bfL\bfxg
\end{equation}

The Jacobian matrix of this system is $\bfJ=\bfGg-\bfL$, where
\[
\bfGg(\bfxg,t)=\left(
\begin{array}{ccc}
\frac{\partial\bff}{\partial\bfx}(\bfx_1,t)&\zeros&\zeros\\
\zeros&\ddots&\zeros\\
\zeros&\zeros&\frac{\partial\bff}{\partial\bfx}(\bfx_n,t)
\end{array}\right)
\]

Let now ($\bfe_1,\dots,\bfe_d$) be a basis of the state space of one
element and consider the following vectors of the global state space
\[
\bfeg_1=\left(
\begin{array}{c}
\bfe_1\\
\vdots\\
\bfe_1
\end{array}\right), 
\dots,
\bfeg_d=\left(
\begin{array}{c}
\bfe_d\\
\vdots\\
\bfe_d
\end{array}\right), 
\]

Let $\sM=\mathrm{span}\{\bfeg_1,\dots, \bfeg_d\}$ be the
``diagonal'' subspace spanned by the $\bfeg_i$. Note that
$\bfxg^\ast\in\sM$ if and only if $\bfx_1^\ast=\dots=\bfx_n^\ast$,
i.e. all elements are in synchrony. In such a case, all coupling
forces equal zero, and the individual dynamics are the same for every
element. Hence
\[
\bffg(\bfxg^\ast,t)-\bfL\bfxg^\ast=
\left(
\begin{array}{c}
\bff(\bfx_1^\ast,t)\\
\vdots\\
\bff(\bfx_1^\ast,t)
\end{array}
\right)
\in\sM
\]
which means that $\sM$ is flow-invariant. 

Consider, as in section \ref{sec:convergence}, the projection matrix
$\bfV$ on $\sM^\perp$. Since $\bfV$ is built from orthonormal vectors,
$\lambda_\mathrm{max}(\bfV\bfGg(\bfxg,t)\bfV^\top)$ is upper-bounded
by $\max_{i}\lambda_\mathrm{max}
\left(\frac{\partial\bff}{\partial\bfx}(\bfx_i,t)\right)$.  Thus, by
virtue of theorem~\ref{theorem:convergence}, a simple sufficient
condition for global exponential synchronization is

\begin{equation}
  \label{equ:synccond}
  \lambda_\mathrm{min}(\bfV\bfL\bfV^\top) \ > \
  \sup_{\bfa,t}\lambda_\mathrm{max}\left(
    \frac{\partial\bff}{\partial\bfx}(\bfa,t)\right)
\end{equation}

Furthermore, the \emph{synchronization rate}, i.e. the rate of
convergence to the synchronization subspace, is the \emph{contraction
  rate} of the auxiliary system (\ref{equ:sysy}).

Let us now make some brief remarks.

\begin{enumerate}

\item \emph{Undirected\,\footnote{``Undirected'' is to be understood
      here in the graph-theoretical sense, i.e. : for all $i,j$, the
      connection from $i$ to $j$ is the same as the one from $j$ to
      $i$. Therefore, an undirected network can be represented by an
      undirected graph, where each edge stands for two connections,
      one in each direction.} diffusive networks.} In this case, it is
  well known that $\bfL$ is symmetric positive semi-definite, and that
  $\sM$ is a subset of the eigenspace corresponding to the
  eigenvalue~0 \cite{Fiedler}. Furthermore, if the network is
  \emph{connected}, this eigenspace is exactly $\sM$, and therefore
  $\bfV\bfL\bfV^\top$ is positive definite (its smallest eigenvalue is
  called the network's \emph{algebraic connectivity} \cite{Fiedler}).
  Assume now that $\bfL$ is parameterized by a positive scalar $k$
  (i.e.  $\bfL=k\bfL_0$, for some $\bfL_0$), and that
  $\frac{\partial\bff}{\partial\bfx}$ is upper-bounded. Then, for
  large enough $k$ (i.e. for strong enough coupling strength), all
  elements will synchronize exponentially.

\item \emph{Network of contracting elements.} If the elements $\bfx_i$
  are already \emph{contracting} when taken in isolation (i.e.
  $\frac{\partial\bff}{\partial\bfx}$ is uniformly
  negative definite), then in presence of \emph{weak or non-existent
    couplings} ($\bfV\bfL\bfV^\top=0$), the Jacobian matrix $\bfJ$
  of the global system will remain uniformly negative definite
  \cite{WangSlo}. Thus, the projected Jacobian matrix will be \emph{a
    fortiori} uniformly negative definite, implying exponential
  convergence to the synchronized state.

  One can also obtain this conclusion by using a ``pure'' contraction
  analysis. Indeed, choose a particular initial state where
  $\bfx_1(0)=\dots=\bfx_n(0)$. The trajectory starting with that
  initial state verifies $\forall t, \bfx_1(t)=\dots=\bfx_n(t)$ by
  flow-invariance. Since the global system is contracting, any other
  initial conditions will lead exponentially to that particular
  trajectory, i.e., starting with any initial conditions, the system
  will exponentially converge to a synchronized state.


\item \emph{Nonlinear couplings.} Similarly to ~\cite{WangSlo}, the
  above result actually extends to nonlinear couplings described by a
  Laplacian matrix $\bfL(\bfxg,t)$. Replacing the auxiliary
  system~(\ref{equ:sysy}) by
\[
\dot{\bfy}=\bfV\bffg(\bfV^\top\bfy+\bfU^\top\bfU\bfxg_p,t)-
\bfV\bfL(\bfxg_p,t)(\bfV^\top\bfy+\bfU^\top\bfU\bfxg_p)
\]
the same steps show that global synchronization is achieved
exponentially for
\[
\inf_{\bfxg,t}
\lambda_\mathrm{min}(\bfV\bfL(\bfxg,t)\bfV^\top) \ > \
\sup_{\bfa,t}\lambda_\mathrm{max}\left(
  \frac{\partial\bff}{\partial\bfx}(\bfa,t)\right)
\]

\item \emph{Leader-followers network.} Assume that there exists a {\it
    leader} $\bfx_\ell$ in the network~\cite{WangSlo}, i.e., an
  element which has no {\it incoming} connections from the other
  elements, $\dot{\bfx}_\ell=\bff(\bfx_\ell,t)$. Convergence to $\sM$
  (guaranteed by satisfying~(\ref{equ:synccond})) then implies that
  all the network elements will synchronize to the leader trajectory
  $\bfx_\ell(t)$.

\item \emph{Non-diffusive couplings.} Note that the above results are
  actually not limited to diffusive couplings but apply to any system
  of the general form~(\ref{equ:syncx}). This point will be further
  illustrated in sections \ref{sec:analysis} and \ref{sec:syminv}.

\end{enumerate}


\section{Main discussion}

In nonlinear contraction theory, the analysis of dynamical systems is
greatly simplified by studying stability and nominal motion
separately. We propose a similar point of view for analyzing
synchronization in networks of dynamical systems. In section
\ref{sec:analysis}, we study specific conditions on the coupling
structure which guarantee exponential convergence to a linear
subspace. In section \ref{sec:syminv}, we examine how symmetries
and/or diffusion-like couplings can give rise to specific
flow-invariant subspaces corresponding to concurrent synchronized
states.

\subsection{Some coupling structures and conditions for exponential
  synchronization}

\label{sec:analysis}

\subsubsection{Balanced diffusive networks}

\label{sec:balanced}

\begin{figure}[ht]
  \begin{minipage}[ht]{0.5\textwidth}
    \centering
    \includegraphics[scale=0.7]{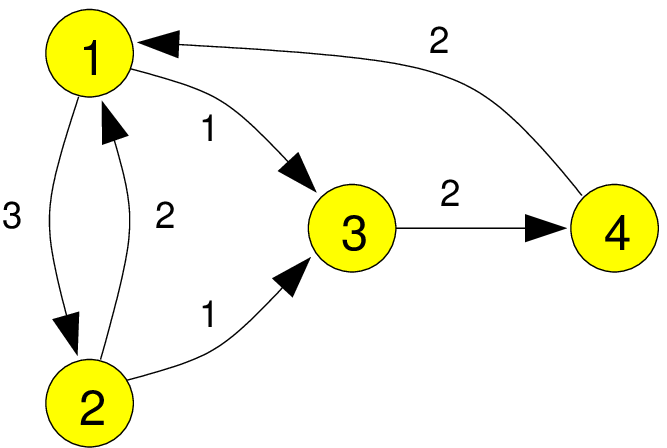}
  \end{minipage}
  \begin{minipage}[ht]{0.5\textwidth}   
    \flushleft
    \caption[]{A balanced network with Laplacian matrix \\
    $\bfL=\left(
        \begin{array}{rrrr}
          4&-2&0&-2\\
          -3&3&0&0\\
          -1&-1&2&0\\
          0&0&-2&2
        \end{array}\right)$}
    \label{fig:balanced}
  \end{minipage} \\[10pt]
  \begin{minipage}[ht]{0.5\textwidth}
    \centering
    \includegraphics[scale=0.7]{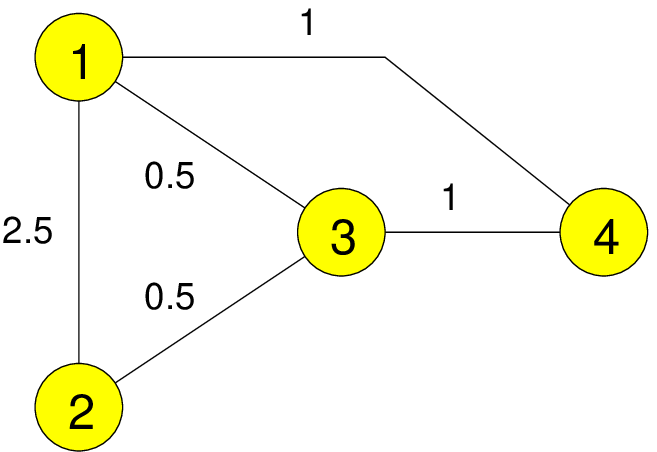}
  \end{minipage}
  \begin{minipage}[ht]{0.5\textwidth}
    \flushleft {Its underlying undirected graph,
      with Laplacian matrix \\
      $\bfL_s=\frac{\bfL+\bfL^\top}{2}=\left(
        \begin{array}{rrrr}
          4&-2.5&-0.5&-1\\
          -2.5&3&-0.5&0\\
          -0.5&-0.5&2&-1\\
          -1&0&-1&2
        \end{array}\right)$} 
  \end{minipage}
\end{figure}

A balanced network \cite{OlfMur} is a directed diffusive network which
verifies the following equality for each node~$i$ (see figure
\ref{fig:balanced} for an example)
\[
\sum_{j \neq i} \bfK_{ij}=
\sum_{j \neq i} \bfK_{ji}
\]

Because of this property, the symmetric part of the Laplacian matrix
of the network is itself the Laplacian matrix of the underlying
undirected graph to the network\,\footnote{In fact, it is easy to see
  that the symmetric part of the Laplacian matrix of a directed graph
  is the Laplacian matrix of some undirected graph \emph{if and only
    if} the directed graph is balanced.}. Thus, the positive
definiteness of $\bfV\bfL\bfV^\top$ for a balanced network is
equivalent to the connectedness of some well-defined undirected graph.

For general directed diffusive networks, finding a simple condition
implying the positive definiteness of $\bfV\bfL\bfV^\top$ (such as the
connectivity condition in the case of undirected networks) still
remains an open problem. However, given a particular example, one can
compute $\bfV\bfL\bfV^\top$ and determine directly whether it is
positive definite.

\subsubsection{Extension of diffusive connections} 

In some applications \cite{WangSlo2}, one might encounter the
following dynamics
\[
\left\{ \begin{array}{l}
\dot{\bfx}_1=\bff_1(\bfx_1,t) + k\bfA^\top(\bfB\bfx_2-\bfA\bfx_1)\\
\dot{\bfx}_2=\bff_2(\bfx_2,t) + k\bfB^\top(\bfA\bfx_1-\bfB\bfx_2)
\end{array} \right.
\]

Here $\bfx_1$ and $\bfx_2$ can be of different dimensions, say $d_1$
and $d_2$. $\bfA$ and $\bfB$ are constant matrices of appropriate
dimensions. The Jacobian matrix of the overall system is
\[
\bfJ=\left( 
\begin{array}{cc}
\frac{\partial\bff_1}{\partial\bfx_1}& \\
& \frac{\partial\bff_2}{\partial\bfx_2}
\end{array} 
\right) - k\bfL, \quad \mathrm{where~} \bfL=
\left(
\begin{array}{cc}
  \bfA^\top\bfA & -\bfA^\top\bfB \\
  -\bfB^\top\bfA & \bfB^\top\bfB
\end{array} 
\right)
\]

Note that $\bfL$ is symmetric positive semi-definite. Indeed, one
immediately verifies that

\[
\forall \bfx_1, \bfx_2 : 
\left(
\begin{array}{cc}
\bfx_1 & \bfx_2
\end{array} 
\right) \bfL \left(
\begin{array}{c}
\bfx_1 \\
\bfx_2
\end{array} 
\right) =
(\bfA\bfx_1-\bfB\bfx_2)^\top (\bfA\bfx_1-\bfB\bfx_2) \geq 0
\]

Consider now the linear subspace of
$\mathbb{R}^{d_1}\times\mathbb{R}^{d_2}$ defined by
\[
\sM=\left\{ \left(
\begin{array}{c}
  \bfx_1 \\
  \bfx_2
\end{array} 
\right) \in \mathbb{R}^{d_1}\times\mathbb{R}^{d_2} :
\bfA\bfx_1-\bfB\bfx_2=\zeros \right\}
\]
and use as before the orthonormal projection $\bfV$ on $\sM^\perp$, so
that $\bfV\bfL\bfV^\top$ is positive definite. Assume furthermore that
$\sM$ is flow-invariant, i.e.
\[
\forall (\bfx_1,\bfx_2)\in \mathbb{R}^{d_1}\times\mathbb{R}^{d_2},
\left[\bfA\bfx_1=\bfB\bfx_2\right] \Rightarrow
\left[\bfA\bff_1(\bfx_1)=\bfB\bff_2(\bfx_2)\right]
\] 
and that the Jacobian matrices of the individual dynamics are
upper-bounded. Then large enough $k$, i.e. for example
\[
k\lambda_\mathrm{min}(\bfV\bfL\bfV^\top)> \max_{i=1,2}\left(
  \sup_{\bfa_i,t}\lambda_\mathrm{max}
  \frac{\partial\bff_i}{\partial\bfx_i}(\bfa_i,t)\right)
\]
ensures exponential convergence to the subspace $\sM$.

The state corresponding to $\sM$ can be viewed as an extension of
synchronization states to systems of different dimensions. Indeed, in
the case where $\bfx_1$ and $\bfx_2$ have the same dimension and where
$\bfA=\bfB$ are non singular, we are in the presence of classical
diffusive connections, which leads us back to the discussion of
section \ref{sec:global}.

As in the case of diffusive connections, one can consider networks of
so-connected elements, for example :
\[
\left\{ \begin{array}{l} \dot{\bfx}_1=\bff_1(\bfx_1,t) +
    \bfA_B^\top(\bfB_A\bfx_2-\bfA_B\bfx_1)
    + \bfA_C^\top(\bfC_A\bfx_3-\bfA_C\bfx_1)\\
    \dot{\bfx}_2=\bff_2(\bfx_2,t) +
    \bfB_C^\top(\bfC_B\bfx_3-\bfB_C\bfx_2)
    + \bfB_A^\top(\bfA_B\bfx_1-\bfB_A\bfx_2)  \\
    \dot{\bfx}_3=\bff_3(\bfx_2,t) +
    \bfC_A^\top(\bfA_C\bfx_1-\bfC_A\bfx_3) +
    \bfC_B^\top(\bfB_C\bfx_2-\bfC_B\bfx_3)
\end{array} \right.
\]
leads to a positive semi-definite Laplacian matrix
\[
\left(
\begin{array}{rrr}
  \bfA_B^\top\bfA_B & -\bfA_B^\top\bfB_A & \zeros\\
  -\bfB_A^\top\bfA_B & \bfB_A^\top\bfB_A & \zeros\\
  \zeros & \zeros & \zeros
\end{array} 
\right)+ \left(
\begin{array}{rrr}
  \zeros & \zeros & \zeros\\
  \zeros & \bfB_C^\top\bfB_C & -\bfB_C^\top\bfC_B \\
  \zeros & -\bfC_B^\top\bfB_C & \bfC_B^\top\bfC_B \\
\end{array} 
\right)+ \left(
\begin{array}{rrr}
  \bfA_C^\top\bfA_C & \zeros & -\bfA_C^\top\bfC_A\\
  \zeros & \zeros & \zeros \\
  -\bfC_A^\top\bfA_C & \zeros & \bfC_A^\top\bfC_A
\end{array} 
\right)
\]
and potentially a flow-invariant subspace
\[
\sM=\{\bfA_B\bfx_1=\bfB_A\bfx_2\} \cap \{\bfB_C\bfx_2=\bfC_B\bfx_3\}
\cap \{\bfC_A\bfx_3=\bfA_C\bfx_1\}
\]

The above coupling structures can be implemented in nonlinear versions
of the predictive hierarchies used in image processing
(e.g.~\cite{LueWil,DayHin,RaoBal,Korner,GeoHaw,Rao}).

\subsubsection{Excitatory-only networks}

\label{sec:excit}

One can also address the case of networks with excitatory-only
connections. Consider for instance the following system and its
Jacobian matrix\,\footnote{For the sake of clarity, the elements are
  assumed to be 1-dimensional. However, the same reasoning applies for
  the multidimensional case as well: instead of $\mathrm{span}{\tiny
    \left\{ \left(\begin{array}{c}
          1\\
          1
      \end{array}\right)
  \right\}}$, one considers
  $\mathrm{span}\{\bfeg_1,\dots,\bfeg_d\}$ as in section
  \ref{sec:global}.}
\[
\left\{\begin{array}{l}
    \dot{x}_1=f(x_1,t)+k x_2 \\
    \dot{x}_2=f(x_2,t)+k x_1
  \end{array}\right.
\qquad \bfJ=\left(\begin{array}{cc}
    \frac{\partial{f}}{\partial{x}}(x_1,t)&0\\
    0&\frac{\partial{f}}{\partial{x}}(x_2,t)
  \end{array}\right) +
k\left(\begin{array}{cc}
    0&1\\
    1&0
  \end{array}\right)
\]

Clearly, $\mathrm{span}\{(1,1)\}$ is flow-invariant. Applying the
methodology described above, we choose
$\bfV=\frac{1}{\sqrt{2}}(1,-1)$, so that the projected Jacobian matrix
is $\frac{1}{2}\left(\frac{\partial{f}}{\partial{x}}(x_1,t)+
  \frac{\partial{f}}{\partial{x}}(x_2,t)\right)-k$.  Thus, for $k >
\sup_{a,t}\frac{\partial{f}}{\partial{x}}(a,t)$, the two elements
synchronize exponentially.

In the case of diffusive connections, once the elements are
synchronized, the coupling terms disappear, so that each individual
element exhibits its natural, uncoupled behavior. This is not the case
with excitatory-only connections. This is illustrated in figure
\ref{fig:fn} using FitzHugh-Nagumo oscillator models (see appendix
\ref{sec:oscillators} for the contraction analysis of coupled
FitzHugh-Nagumo oscillators).

\begin{figure}[ht]
  \begin{minipage}[ht]{0.32\textwidth}
    \centering
    \includegraphics[scale=0.35]{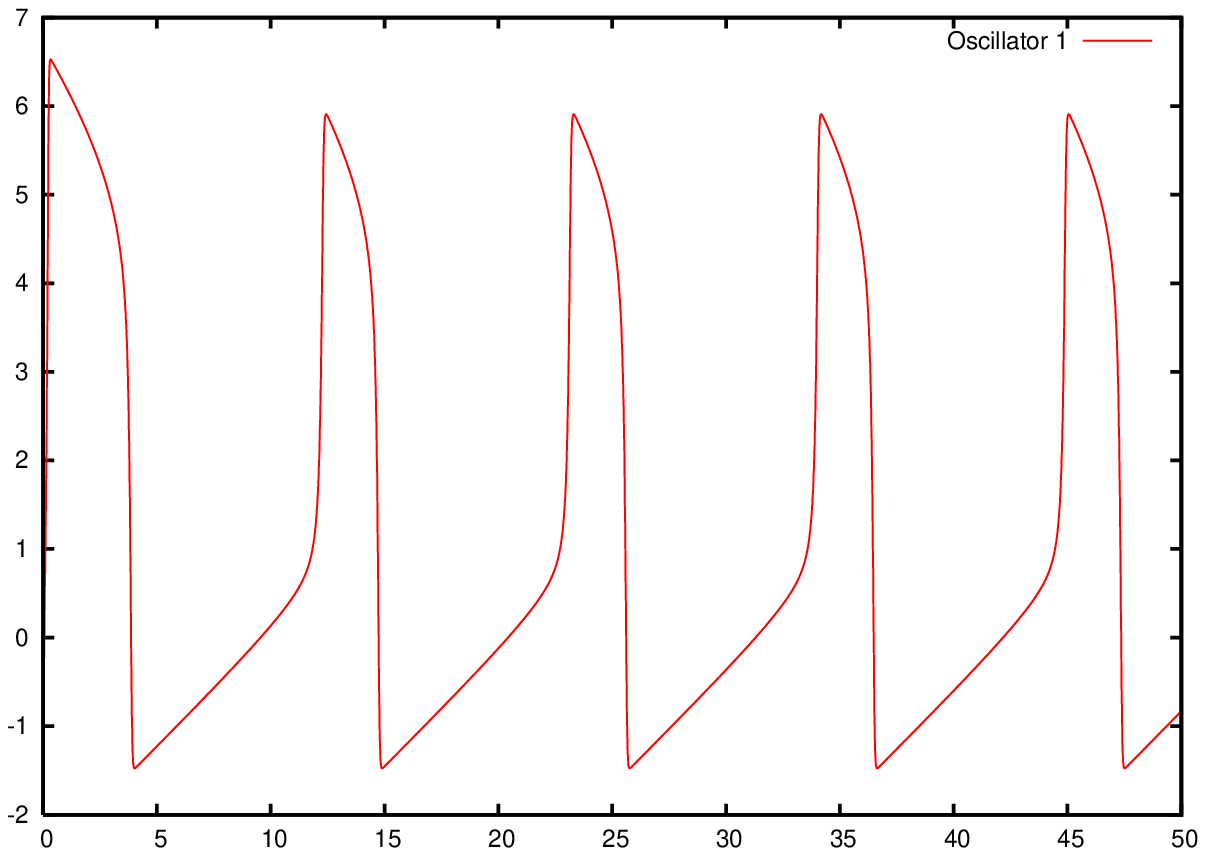}
  \end{minipage}
  \begin{minipage}[ht]{0.32\textwidth}
    \centering
    \includegraphics[scale=0.35]{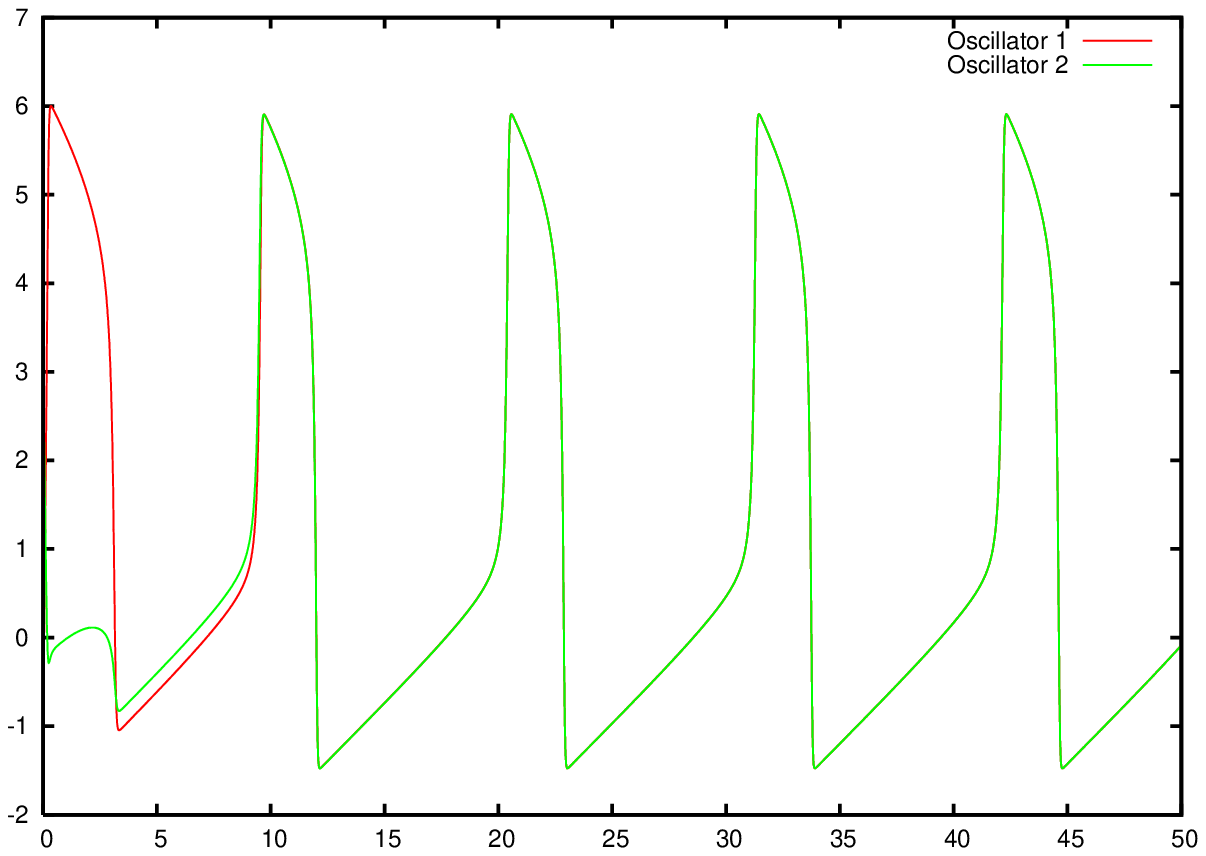}
  \end{minipage}
  \begin{minipage}[ht]{0.32\textwidth}
    \centering
    \includegraphics[scale=0.35]{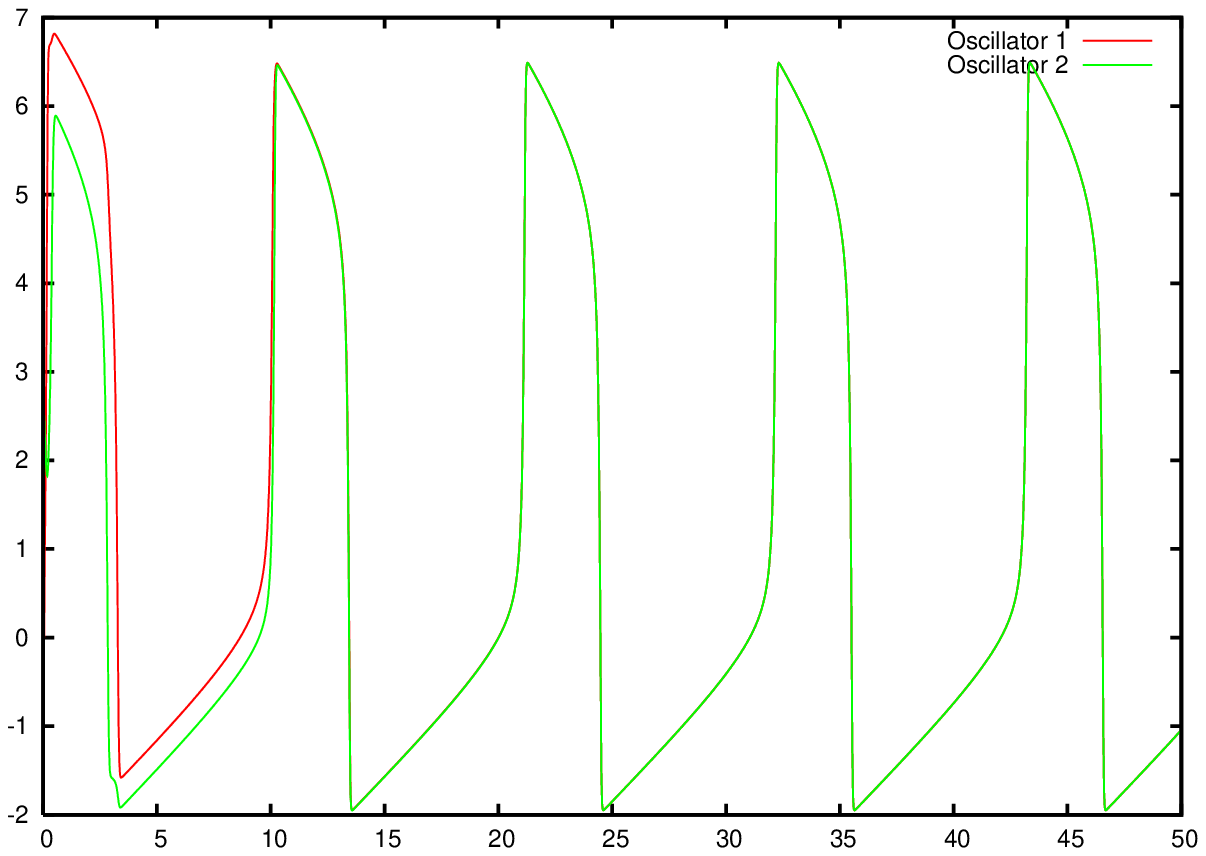}
  \end{minipage}
  \caption[]{From left to right : a single oscillator, two
    oscillators coupled through diffusive connections, two oscillators
    coupled through excitatory-only connections.}
  \label{fig:fn}
\end{figure}

\subsubsection{Rate models for neuronal populations} 

\label{sec:rate}

In computational neuroscience, one often uses the following simplified
equations to model the dynamics of neuronal populations
\[
\tau \dot{\bfx}_i= -\bfx_i+ \Phi\left(\sum_{j\neq
    i}k_{ij}\bfx_j(t)\right)+ \bfu_i(t)
\]

Assume that the external inputs $\bfu_i(t)$ are all equal, and that
the synaptic connections $k_{ij}$ verify $ \exists c, \forall i,\
\sum_{j\neq i}k_{ij} = c $ (i.e., that they induce input-equivalence, see
section~\ref{sec:syminv}). Then the synchronization subspace
\mbox{$\{\bfx_1=\dots=\bfx_n\}$} is flow-invariant. Furthermore, since
each element, taken in isolation, is \emph{contracting} with
contraction rate $1/\tau $, synchronization should occur when the
coupling is not too strong (see remark (ii) in
section~\ref{sec:global}).

Specifically, consider first the case where $\Phi$ is a linear
function : $\Phi(\bfx)=\mu\bfx$. The Jacobian matrix of the global
system is then $-\bfI_n+\mu\bfK$, where $\bfK$ is the matrix of
$k_{ij}$.  Using the result of remark (ii) in section
\ref{sec:global}, a sufficient condition for the system to be
contracting (and thus synchronizing) is that the couplings are
\emph{weak} enough (or more precisely, such that
$\mu\lambda_\mathrm{max}(\bfK) < 1$).

The same condition is obtained if $\Phi$ is now e.g. a
multidimentional \emph{sigmoid} of maximum slope $\mu$ (see remark
(iii) in section \ref{sec:global}).

Besides the synchronization behavior of these models, their natural
\emph{contraction} property for weak enough couplings of any sign is
interesting in its own right. Indeed, given a set of (not necessarily
equal) external inputs $\bfu_i(t)$, all trajectories of the global
system will converge to a unique trajectory, independently of initial
conditions.

\subsection{Symmetries, diffusion-like couplings, flow-invariant
  subspaces and concurrent synchronization}

\label{sec:syminv}

Synchronized states can be created in at least two ways : by
architectural and internal\footnote{Internal symmetries can easily
be analyzed within our framework as leading to flow-invariant
subspaces, and we shall use this property in section \ref{sec:cpg} for
building central pattern generators. However, they will not be
discussed in detail in this article. The interested reader can consult
\cite{DioGolSte}.}  symmetries
\cite{GolSte,GolSteTor,DioGolSte,PogSanNij} or by diffusion-like
couplings \cite{WangSlo,JadMotBar,OlfMur,LinBroFra,Belykh1}. Actually,
we shall see that both, together or separately, can create
flow-invariant subspaces corresponding to concurrently synchronized
states.

\subsubsection{Symmetries and input-equivalence}

In section \ref{sec:global}, we argued that, in the case of coupled
\emph{identical} elements, the global synchronization subspace $\sM$
represents a flow-invariant linear subspace of the global state space.
However, several previous works have pointed out that larger (less
restrictive) flow-invariant subspaces may exist if the network
exhibits symmetries~\cite{Zhang,Belykh1,PogSanNij}, even when the
systems are {\it not} identical~\cite{GolSte}.

The main idea behind these works can be summarized as follows. Assume
that the network is divided into $k$ aspiring synchronized groups
$S_1,\dots,S_k$\,\footnote{Some groups may contain a single element,
 see section \ref{sec:single}.}. The flow-invariant subspace
corresponding to this regime (in the sequel, we shall call such a
subspace a \emph{concurrent synchronization subspace}), namely
\[
\{(\bfx_1;\dots;\bfx_n) : \forall 1\leq m \leq k,
\forall i,j \in S_m : \bfx_i=\bfx_j\}
\]
is flow-invariant if, for each $S_m$, the following conditions are
true :

\begin{enumerate}
\item if $i,j\in S_m$, then they have a same individual (uncoupled)
  dynamics
\item if $i,j\in S_m$, and if they receive their input from elements
  $i'$ and $j'$ respectively, then $i'$ and $j'$ must be in a same
  group $S_{m'}$, and the coupling functions (the synapses)
  $i'\rightarrow i$ and $j'\rightarrow j$ must be identical. If $i$
  and $j$ have more than one input, they must have the same number of
  inputs, and the above conditions must be true for each input.  In
  this case, we say that $i$ and $j$ are input-symmetric, or more
  precisely, {\it input-equivalent} (since formally ``symmetry'' implies
  the action of a group).
\end{enumerate}

One can see here that symmetry, or more generally
input-equivalence, plays a key role in concurrent
synchronization.  For a more detailed discussion, the reader is
referred to \cite{GolSte,GolSteTor}.

\textbf{Remark :} One can thus turn on/off a specific symmetry by
turning on/off a single connection. This has similarities to the fact
that a single inhibitory connection can turn on/off an entire network
of synchronized identical oscillators~\cite{WangSlo}.

\subsubsection{Diffusion-like couplings} 

The condition of input-equivalence can be relaxed when some
connections \emph{within a group} are null when the connected elements
are in the same state. Such connections are pervasive in the
literature : diffusive connections (in a neuronal context, they
correspond to electrical synapses mediated by gap junctions
\cite{SheRin,fukuda06}, in an automatic control context, they
correspond to poursuit or velocity matching strategies
\cite{OlfMur,LinBroFra}, \dots), connections in the Kuramoto model
\cite{IzhiKura,JadMotBar,Strogatz} (i.e. in the form
$\dot{x_i}=f(x_i,t)+\sum_{j}k_{ij}\sin(x_j-x_i)$), etc.

Indeed, consider for instance diffusive connections and assume that
\begin{itemize}
\item $i,i',j,j' \in S_m$
\item $i'\rightarrow i$ has the form $\bfK_1(\bfx_{i'}-\bfx_i)$
\item $j'\rightarrow j$ has the form $\bfK_2(\bfx_{j'}-\bfx_j)$ with
  possibly $\bfK_1\neq \bfK_2$
\end{itemize}
Here, $i$ and $j$ are not input-equivalent in the sense we defined
above, but the subspace $\{\bfx_i=\bfx_j=\bfx_{i'}=\bfx_{j'}\}$ is
still flow-invariant. Indeed, once the system is on this
synchronization subspace, we have $\bfx_i=\bfx_{i'}$,
$\bfx_j=\bfx_{j'}$, so that the diffusive couplings $i'\rightarrow i$
and $j'\rightarrow j$ vanish.

One can also view the network as a directed graph $G$, where the
elements are represented by nodes, and connections $i\to j$ by
directed arcs $i\to j$. Then, the above remark can be reformulated as

\begin{description}
\item[1 :] for all $m$, color the nodes of $S_m$ with a color $m$,
\item[2 :] for all $m$, erase the arcs representing diffusion-like
  connections and joining two nodes in $S_m$,
\item[3 :] check whether the initial coloring is balanced (in the
  sense of \cite{GolSte}) with respect to the so-obtained graph.
\end{description}

It should be clear by now that our framework is particularly suited to
analyze concurrent synchronization. Indeed, a general methodology to
show global exponential convergence to a concurrent synchronization
regime consists in the following two steps

\begin{itemize}
\item First, find an flow-invariant linear subspace by taking advantage of
  potential symmetries in the network and/or diffusion-like connections.
\item Second, compute the projected Jacobian matrix on the orthogonal
  subspace and show that it is uniformly negative definite (by
  explicitly computing its eigenvalues or by using results regarding
  the form of the network, e.g. remark (i) in section \ref{sec:global}
  or section \ref{sec:analysis}).
\end{itemize}

\subsection{Illustrative examples} 

\begin{figure}[ht]
  \begin{minipage}[ht]{0.31\textwidth}
    \centering
    \includegraphics[scale=0.4]{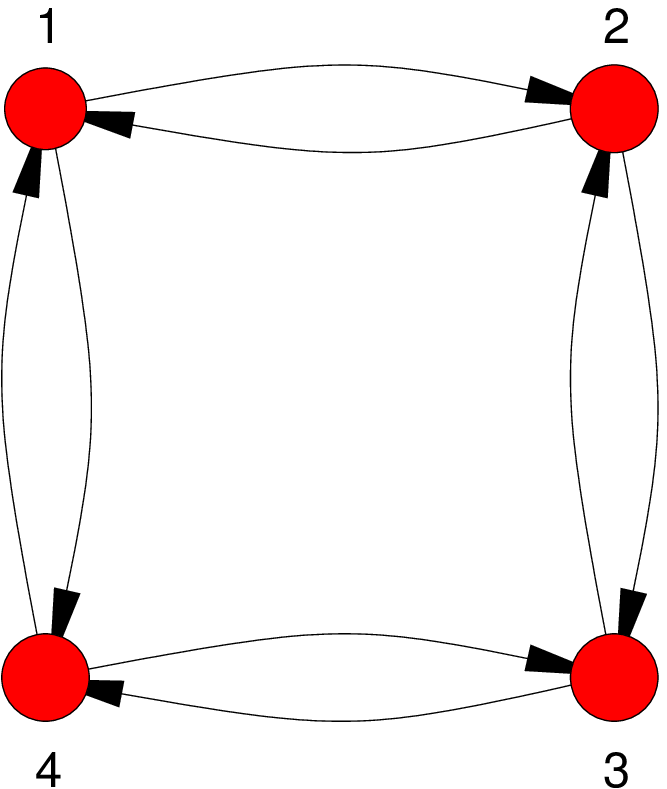}
  \end{minipage}
  \begin{minipage}[ht]{0.31\textwidth}
    \centering
    \includegraphics[scale=0.4]{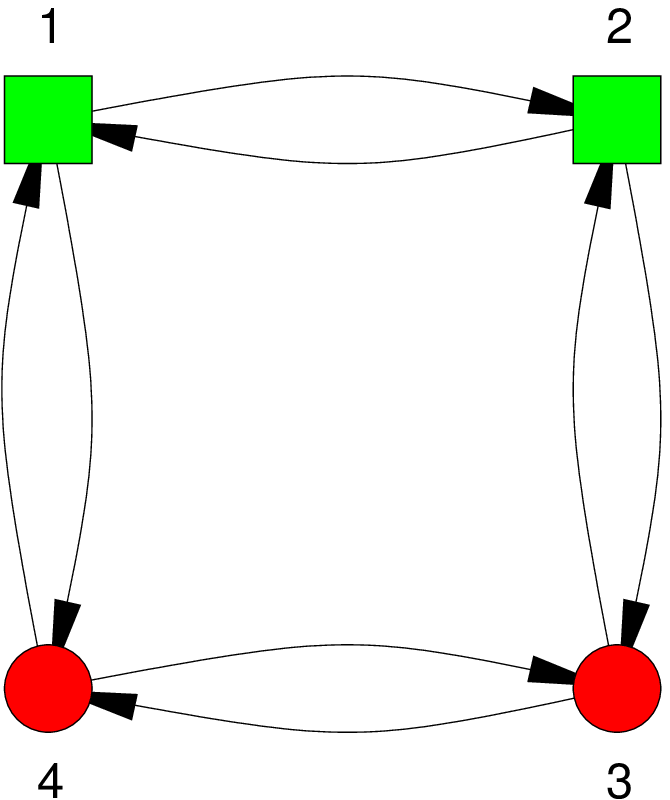}
  \end{minipage}
  \begin{minipage}[ht]{0.33\textwidth}
    \centering
    \includegraphics[scale=0.45]{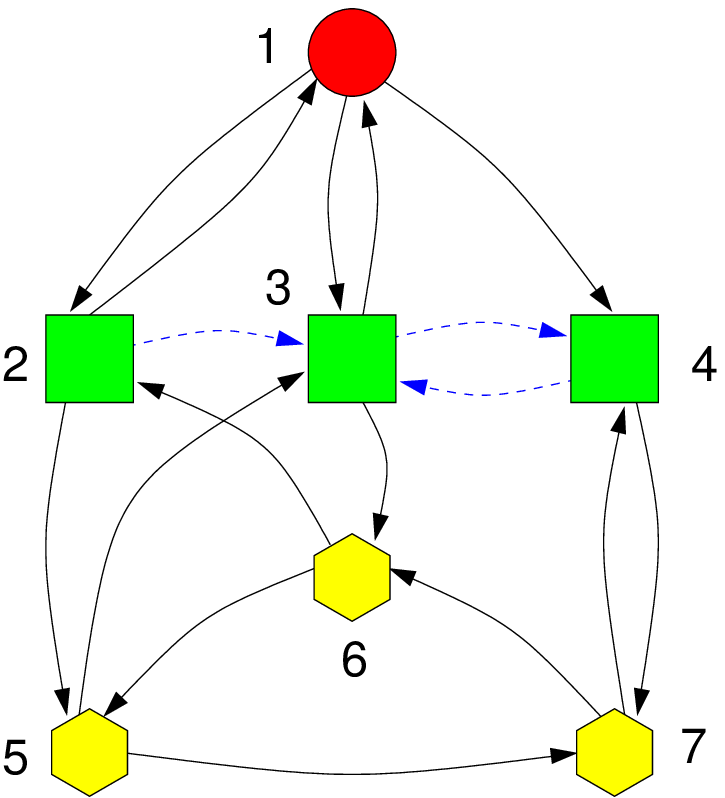}
  \end{minipage}
  \caption[]{Three example networks}
  \label{fig:examples}
\end{figure}

\begin{enumerate}
\item The first network has three non-trivial flow-invariant subspaces
  other than the global sync subspace, namely $\sM_1=\{\bfx_1=\bfx_2,
  \bfx_3=\bfx_4\}$, $\sM_2=\{\bfx_1=\bfx_3, \bfx_2=\bfx_4\}$, and
  $\sM_3=\{\bfx_1=\bfx_4, \bfx_2=\bfx_3\}$.  Any of these subspaces is
  a strict superset of the global sync subspace, and therefore one
  should expect that the convergence to any of the concurrent sync
  state is ``easier'' than the convergence to the global sync
  state~\cite{Zhang,Belykh1,PogSanNij}. This can be quantified from
  (\ref{equ:synccond}), by noticing that 
\begin{equation} \label{percolation}
\sM_A \supset \sM_B 
  \Rightarrow\sM_A^\perp \subset \sM_B^\perp \ \Rightarrow
  \lambda_\mathrm{min}(\bfV_A\bfL\bfV_A^\top) \ge
  \lambda_\mathrm{min}(\bfV_B\bfL\bfV_B^\top) \ \ \
\end{equation}
While in the case of identical systems and relatively uniform
topologies, this ``percolation'' effect may often be too fast to
observe, (\ref{percolation}) applies to the general concurrent
synchronization case and quantifies the associated and possibly very
distinct time-scales.
\item The second network has only one non-trivial flow-invariant
  subspace $\{\bfx_1=\bfx_2, \bfx_3=\bfx_4\}$.
\item If the dashed blue arrows represent diffusive connections then
  the third network will have one non-trivial flow-invariant subspace
  $\{\bfx_2 =\bfx_3=\bfx_4, \bfx_5=\bfx_6=\bfx_7\}$, even if these
  extra diffusive connections \emph{obviously break the symmetry}.
\end{enumerate}

\begin{figure}[ht]
  \centering
  \includegraphics[scale=0.5]{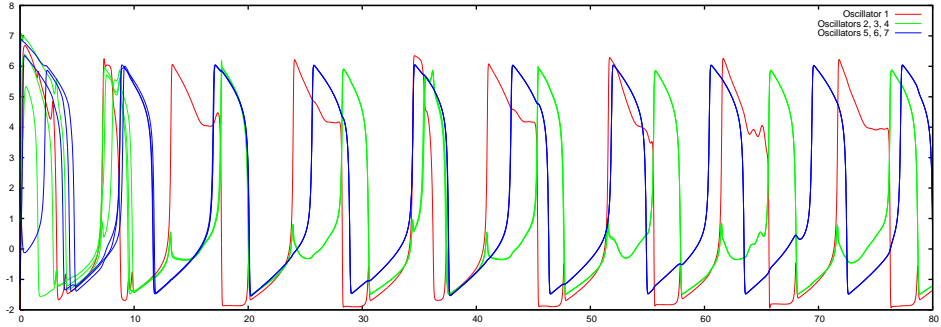}
  \caption{Simulation result for network 3.}
  \label{fig:exfig}
\end{figure}

Let's study in more detail this third network, in which the
connections between the round element and the square ones are modelled
by trigonometric functions (we shall see in section \ref{sec:single}
that their exact form has no actual influence on the convergence
rate).
\[ 
\left\{ \begin{array}{l}
    \dot{v}_1=f(v_1)+a_1\cos(v_2)+a_2\sin(v_3)\\
    \dot{v}_2=g(v_2)+a_4\sin(v_1)+c_1v_6\\
    \dot{v}_3=g(v_3)+a_4\sin(v_1)+b_1(v_2-v_3)+b_2(v_4-v_3)+c_1v_5\\
    \dot{v}_4=g(v_4)+a_4\sin(v_1)+b_3(v_3-v_4)+c_1v_7 \\
    \dot{v}_5=h(v_5)+c_2v_2+(d_2v_7-d_1v_5)\\
    \dot{v}_6=h(v_6)+c_2v_3+(d_2v_5-d_1v_6)\\
    \dot{v}_7=h(v_7)+c_2v_4+(d_2v_6-d_1v_7)\\
  \end{array} \right.  \] 
The Jacobian matrix of the couplings is 
\[
\bfL=\left( \begin{array}{rrrrrrr}
    0&a_1\dot{v}_2\sin(v_2)&-a_2\dot{v}_3\cos(v_3)&0&0&0&0 \\
    -a_4\dot{v}_1\cos(v_1)&0&0&0&0&-c_1&0 \\
    -a_4\dot{v}_1\cos(v_1)&-b_1&b_1+b_2&-b_2&-c_1&0&0 \\
    -a_4\dot{v}_1\cos(v_1)&0&-b_3&b_3&0&0&-c_1 \\
    0&-c_2&0&0&d_1&0&-d_2 \\
    0&0&-c_2&0&-d_2&d_1&0 \\
    0&0&0&-c_2&0&-d_2&d_1 \end{array}\right) 
\] 
As we remarked previously, the concurrent synchronization regime
$\{v_2=v_3=v_4,v_5=v_6=v_7\}$ is possible. Bases of the linear
subspaces $\sM$ and $\sM^\perp$ corresponding to this regime are \[
\left(\begin{array}{r} 1\\0\\0\\0\\0\\0\\0 \end{array} \right),
\left(\begin{array}{r} 0\\1\\1\\1\\0\\0\\0 \end{array} \right),
\left(\begin{array}{r} 0\\0\\0\\0\\1\\1\\1 \end{array} \right) \quad
\textrm{for } \sM,\textrm{ and}\quad \left(\begin{array}{r}
    0\\\frac{\sqrt6}{3}\\\frac{-\sqrt6}{6}\\\frac{-\sqrt6}{6}\\0\\0\\0
  \end{array} \right), \left(\begin{array}{r}
    0\\0\\\frac{-\sqrt2}{2}\\\frac{\sqrt2}{2}\\0\\0\\0 \end{array}
\right), \left(\begin{array}{r}
    0\\0\\0\\0\\\frac{\sqrt6}{3}\\\frac{-\sqrt6}{6}\\\frac{-\sqrt6}{6}
  \end{array} \right), \left(\begin{array}{r}
    0\\0\\0\\0\\0\\\frac{-\sqrt2}{2}\\\frac{\sqrt2}{2} \end{array}
\right) \quad \textrm{for }\sM^\perp.  \] Group together the vectors
of the basis of $\sM^\perp$ into a matrix $\bfV$ and compute
\[
\bfV\bfL_s\bfV^\top=\left(\begin{array}{rrrr}
    \frac{b_1}{2}&-\frac{\sqrt{3}(2b_1+b_2-b_3)}{6}&\frac{c_1-2c_2}{4}&-\frac{c_1\sqrt{3}}{4}\\
    -\frac{\sqrt{3}(2b_1+b_2-b_3)}{6}&\frac{b_1+2(b_2+b_3)}{2}&-\frac{c_1\sqrt{3}}{4}&-\frac{c_1+2c_2}{4}\\
    \frac{c_1-2c_2}{4}&-\frac{c_1\sqrt{3}}{4}&\frac{2d_1+d_2}{2}&0\\
    -\frac{c_1\sqrt{3}}{4}&-\frac{c_1+2c_2}{4}&0&\frac{2d_1+d_2}{2}
    \end{array}\right) \] As a numerical example, let $b_1=3\alpha$,
  $b_2=4\alpha$, $b_3=5\alpha$, $c_1=\alpha$, $c_2=2\alpha$,
  $d_1=3\alpha$, $d_2=4\alpha$ and evaluate the eigenvalues of
  $\bfV\bfL_s\bfV^\top$.  We obtain approximately $1.0077\alpha$ for
  the smallest eigenvalue.  Using again FitzHugh-Nagumo oscillators
  and based on their contraction analysis in appendix
  \ref{sec:oscillators}, concurrent synchronization should occur for
  $\alpha>10.25$.  A simulation is shown in figure \ref{fig:exfig}.
  One can see clearly that, after a transient period, oscillators 2,
  3, 4 are in perfect sync, as well as oscillators 5, 6, 7, but that
  the two groups are not in sync with each other.

\subsection{Robustness of synchronization}

So far, we have been considering \emph{exact} synchronization of
\emph{identical} elements. However this assumption may seem
unrealistic, since real systems are never absolutely identical.  We
use here the robustness result for contracting systems (see
theorem~\ref{theorem:robust}) to guarantee approximate synchronization
even when the elements are not identical.

Consider, as in section \ref{sec:global}, a network of
$n$ dynamical elements
\begin{equation}
  \label{equ:robustsync}
  \dot{\bfx}_i = \bff_i(\bfx_i,t) + \sum_{j \neq i} \bfK_{ij}(\bfx_j -
  \bfx_i) \qquad i=1,\ldots,n
\end{equation}
with now possibly $\bff_i\neq\bff_j$ for $i\neq j$. This can be
rewritten as
\begin{equation}
  \label{equ:robust}
  \left(\begin{array}{c}
      \dot{\bfx}_1\\
      \vdots\\
      \dot{\bfx}_n
    \end{array}\right) =
  \left(\begin{array}{c}
      \bfc(\bfx_1,t)\\
      \vdots\\
      \bfc(\bfx_n,t) 
   \end{array}\right) - 
 \bfL
 \left(\begin{array}{c}
     \bfx_1\\
     \vdots\\
     \bfx_n
  \end{array}\right)+
  \left(\begin{array}{c}
    \bff_1(\bfx_1,t)-\bfc(\bfx_1,t)\\
    \vdots\\
    \bff_n(\bfx_n,t)-\bfc(\bfx_n,t)
  \end{array}\right)
\end{equation}
where $\bfc$ is some function to be defined later. Keeping the
notations introduced in section \ref{sec:global}, one has 
\[
\dot{\bfxg}=\bfcg(\bfxg,t)-\bfL\bfxg+\bfd(\bfxg,t) 
\]
where $\bfd(\bfxg,t)$ stands for the last term of equation
(\ref{equ:robust}).

Consider now the projected auxiliary system on $\sM^\perp$
\begin{equation}
  \label{equ:projecteddisturbed}
  \dot{\bfy}=\bfV\bfcg(\bfV^\top\bfy+\bfU\bfU^\top\bfxg,t)-
  \bfV\bfL\bfV^\top\bfy+\bfV\bfd(\bfV^\top\bfy+\bfU\bfU^\top\bfxg,t)
\end{equation}


Assume that the connections represented by $\bfL$ are strong enough
(in the sense of equation (\ref{equ:synccond})), so that the
undisturbed version of (\ref{equ:projecteddisturbed}) is contracting
with rate $\lambda>0$. Let $D=\sup_{\bfxg,t}\|\bfV\bfd(\bfxg,t)\| $,
where $D$ can be viewed as a measure of the dissimilarity of the
elements.  Since $\bfy=\zeros$ is a particular solution of the
undisturbed system, theorem \ref{theorem:robust} implies that the
distance $R(t)$ between any trajectory of
(\ref{equ:projecteddisturbed}) and $\zeros$ verifies, after a
transient period, $R(t)\leq D/\lambda$. In the $\bfx$-space, it means
that \emph{any trajectory will eventually be contained in a boundary
  layer of thickness $D/\lambda$ around the synchronization subspace
  $\sM$}.

The choice of $\bfc$ can now be specified so as to minimize
$D/\lambda$. Neglecting for simplicity the variation of $\lambda$, a
possible choice for $\bfc(\bfx,t)$ is then the center of the ball of
smallest radius containing $\bff_1(\bfx,t),\dots,\bff_n(\bfx,t)$, with
$D$ being the radius of that ball.

Consider for instance, the following system (similar to the model used
for coincidence detection in \cite{WangSlo} and section
\ref{sec:manies})
\[
\dot{x}_i=f(x_i)+ I_i + k(x_0-x_i) \qquad {\rm where}\ I_\mathrm{min}
\le I_i \le I_\mathrm{max},\ \forall i
\]
In this case, choosing
$c(x)=f(x)+\frac{I_\mathrm{max}+I_\mathrm{min}}{2}$, one can achieve
the bound $D/\lambda$, where $\lambda$ is the contraction rate of $f$
and $D=\frac{I_\mathrm{max}-I_\mathrm{min}}{2}$.

\textbf{Remark :} Assume that two spiking neurons are approximately
synchronized, as just discussed. Then, since spiking induces large
abrupt variations, the neurons must spike approximately at the same
time. More specifically, if the bound on their trajectory discrepancy
guaranteed by the above robustness result is significantly smaller
than spike size, then this bound will automatically imply that the two
neurons spike approximately at the same time.


\section{Combinations of concurrently synchronized groups}
 
This section shows that, under mild conditions, global convergence to
a concurrently synchronized regime is preserved under basic system
combinations, and thus that stable concurrently synchronized
aggregates of arbitrary size can be systematically constructed. The
results, derived for two groups, extend by recursion to arbitrary
numbers of groups.

\label{sec:combination}

\subsection{The input-equivalence preservation assumption}

\label{sec:input-equivalence}

Consider two independent groups of dynamical elements, say $S_1$ and
$S_2$. For each group $i$ ($i=1,2$), assume that a flow-invariant
subspace $\sM_i$ corresponding to a concurrently synchronized regime
exists. Assume furthermore that contraction to this subspace can be
shown, i.e.  $\bfV_i \bfJ_i \bfV_i^\top<\zeros$ for some
projection matrix $\bfV_i$ on $\sM_i^\perp$.

Connect now the elements of $S_1$ to the elements of $S_2$,
\emph{while preserving input-equivalence} for each aspiring synchronized
subgroup of $S_1$ and $S_2$. Thus, the combined concurrent
synchronization subspace $\sM_1\times\sM_2$ remains a flow-invariant
subspace of the new global space. A projection matrix on
$\left(\sM_1\times\sM_2\right)^\perp$ can be $\bfV=\left(
\begin{array}{cc}
\bfW_1 & \zeros \\
\zeros & \bfW_2
\end{array} 
\right)$ where each $\bfV_i$ has been rescaled into $\bfW_i$ to
preserve orthonormality.

Specific mechanisms facilitating input-equivalence preservation will be
discussed in section \ref{sec:single}.

\subsection{Typology of combinations}

\label{sec:typology}

Let us now study several combination operations of concurrently
synchronized groups and discuss how they can preserve convergence to a
combined concurrent sync state.

In \ref{sec:negfeedback} and \ref{sec:hierarchy}, the input-equivalence
preservation condition of section \ref{sec:input-equivalence} is
implicitly assumed, and the results reflect similar combination
properties of contracting systems \cite{LohSlo,Slo,TabSlo}. More
generally, as long as input-equivalence is preserved, any combination
property for contracting systems can be easily ``translated'' into a
combination property for synchronizing systems.

\subsubsection{Negative feedback combination} 
\label{sec:negfeedback}

The Jacobian matrices of the couplings are of the form
$\bfJ_{12}=-k\bfJ_{21}^\top$, with $k$ a positive constant. Thus, the
Jacobian matrix of the global system can be written as 
\[\bfJ=\left(
\begin{array}{cc}
\bfJ_1 & -k\bfJ_{21}^\top \\
\bfJ_{21} & \bfJ_2
\end{array}
\right)
\]
As in equation~(\ref{metric}) of section~\ref{sec:convergence},
consider a transform $\bfTh$ over $ (\sM_1\times\sM_2)^\perp$
\[
\bfTh=\left(
\begin{array}{cc}
\bfI & \zeros \\
\zeros & \sqrt{k}\bfI
\end{array}
\right)
\]

The corresponding \emph{generalized} projected Jacobian matrix on
$(\sM_1\times\sM_2)^\perp$ is
\[
\bfTh(\bfV\bfJ\bfV^\top)\bfTh^{-1}=
\left(
  \begin{array}{cc}
    \bfW_1\bfJ_1{\bfW_1}^\top & \frac{1}{\sqrt{k}}\bfW_1(-k\bfJ_{21}^\top){\bfW_
      2}^\top \\
    \sqrt{k}\bfW_2\bfJ_{21}{\bfW_1}^\top &
    \bfW_2\bfJ_2{\bfW_2}^\top 
  \end{array}
\right) < \zeros \quad {\rm uniformly}
\]
so that global exponential convergence to the combined concurrent
synchronization state can be then concluded.

\subsubsection{Hierarchical combination}
\label{sec:hierarchy}

Assume that the elements in $S_1$ provide feedforward to elements in
$S_2$ but do not receive any feedback from them. Thus, the Jacobian
matrix of the global system is $\bfJ=\left(
\begin{array}{cc}
  \bfJ_1 & \zeros \\
  \bfJ_{21} & \bfJ_2
\end{array}
\right)$. Assume now that $\bfW_2\bfJ_{21}{\bfW_1}^\top$ is uniformly
bounded and consider the coordinate transform $\bfTh_\epsilon=\left(
\begin{array}{cc}
\bfI & \zeros \\
\zeros & \epsilon\bfI
\end{array}
\right)$. Compute the generalized projected Jacobian
\[
\bfTh_\epsilon(\bfV\bfJ\bfV^\top)\bfTh_\epsilon^{-1}= \left(
\begin{array}{cc}
  \bfW_1\bfJ_1{\bfW_1}^\top & \zeros \\
  \epsilon\bfW_2\bfJ_{21}{\bfW_1}^\top & \bfW_2\bfJ_2{\bfW_2}^\top 
\end{array}
\right)
\]
Since $\bfW_2\bfJ_{21}{\bfW_1}^\top$ is bounded, and
$\bfW_1(\bfJ_1){\bfW_1}^\top$ and $\bfW_2(\bfJ_2){\bfW_2}^\top$
are both negative definite,
$\bfTh_\epsilon(\bfV\bfJ\bfV^\top)\bfTh_\epsilon^{-1}$
will be negative definite for small enough $\epsilon$.

Note that classical graph algorithms~\cite{Knu} allow large system
aggregates to be systematically decomposed into hierarchies (directed
acyclic graphs, feedforward networks) of simpler subsystems (strongly
connected components)~\cite{TabSlo}. Input-equivalence then needs only be
verified top-down.

\subsubsection{Case where $S_1$ has a single element} 
\label{sec:single}

Denote this element by $e_1$ (figure \ref{fig:thalamus} shows such a
configuration where $e_1$ is the round red central element, and where
$S_2$ is the set of the remaining elements). Connections from (resp.
to) $e_1$ will be called $1{\small \to}2$ (resp. $2{\small \to}1$)
connections. Then some simplifications can be made :

\begin{itemize}
\item Input-equivalence is preserved whenever, for each aspiring
  synchronized subgroup of $S_2$, the $1{\small \to}2$ connections are
  identical for each element of this subgroup (in figure
  \ref{fig:thalamus}, the connections from $e_1$ to the yellow diamond
  elements are the same). In particular, one can add/suppress/modify
  any $2{\small \to}1$ connection without altering input-equivalence.
\item Since $\mathrm{dim}(\sM_1^\perp)=0$ (a single element is always
  synchronized with itself), one has
  $(\sM_1\times\sM_2)^\perp=\sM_2^\perp$. Thus, concurrent
  synchronization (and the rate of convergence) of the combined system
  only depends on the parameters and the states of the elements within
  $S_2$. In particular, it neither depends on the actual state of
  $e_1$, nor on the connections from/to $e_1$ (in figure
  \ref{fig:thalamus}, the black arrows towards the red central element
  are arbitrary).
\end{itemize}

In practice, the condition of identical $1{\small \to}2$ connections
is quite pervasive. In a neuronal context, one neuron with high
fan-out may have $10^4$ identical outgoing connections.

It is therefore quite easy to preserve an existing concurrent
synchonization behavior while adding groups consisting of a single
element. Thus, stable concurrent synchronization can be easily built
one element at a time.

\subsubsection{The diffusive case} 
\label{sec:diffusive}

We stick with the case where $S_1$ consists of a single element, but
make now the additionnal requirement that the $1{\small \to}2$
connections and the internal connections within $S_2$ are
\emph{diffusive} (so far in this section \ref{sec:combination}, we
have implicitly assumed that the connections from $S_i$ to $S_j$ only
involve the states of elements in $S_i$). The Jacobian matrix of the
combined system is now of the form
\[
\bfJ=\left(
\begin{array}{cccc}
  \frac{\partial g(x_1,t)}{\partial x_1}&&&\\
  &\frac{\partial f_1(x_2,t)}{\partial x_2}&&\\
  &&\ddots&\\
  &&&\frac{\partial f_q(x_n,t)}{\partial x_n}
\end{array}\right)+
\left(\begin{array}{cccc}
    *&*&*&*\\
    k_1   &-k_1&&\\
    \vdots&&\ddots&\\
    k_q &&&-k_q
\end{array}\right)-
\left(\begin{array}{cc}
    0&0\\
    0&\bfL_\mathrm{int}\\
\end{array}\right)
\]
where the first matrix describes the internal dynamics of each
element, the second, the diffusive connections between $e_1$ and $S_2$
(where $S_2$ has $q$ aspiring synchronized subgroups), and the third,
the internal diffusive connections within $S_2$.

Hence, the projected Jacobian matrix on
$(\sM_1\times\sM_2)^\perp=\sM_2^\perp$ is
\[
\bfV_2 \left(\begin{array}{ccc}
    \frac{\partial f_1(x_2,t)}{\partial x_2}&&\\
    &\ddots&\\
    &&\frac{\partial f_q(x_n,t)}{\partial x_n}
\end{array}\right)
\bfV_2^\top- \bfV_2 \left(\begin{array}{ccc}
    k_1&&\\
    &\ddots&\\
    &&k_q
\end{array}\right)
\bfV_2^\top
 - \bfV_2\bfL_\mathrm{int}\bfV_2^\top
\]


An interpretation of this remark is that there are basically three
ways to achieve concurrent synchronization within $S_2$,
\emph{regardless of the behavior of element $e_1$ and of its
  connections} :

\begin{enumerate}
\item one can increase the strengths $k_1,\dots,k_q$ of the $1{\small
    \to}2$ connections (which corresponds to adding inhibitory damping
  to $S_2$), so that each element of $S_2$ becomes \emph{contracting}.
  In this case, all these elements will synchronize because of their
  contracting property even \emph{without any direct coupling} among
  them ($\bfL_\mathrm{int}=0$) (this possibility of synchronization
  without direct coupling is exploited in the coincidence detection
  algorithm of \cite{WangSlo5}, and again in section \ref{sec:manies}
  of this paper),
\item or one can increase the strength $\bfL_\mathrm{int}$ of the
  internal connections among the elements of $S_2$,
\item or one can combine the two.
\end{enumerate}

\subsubsection{Parallel combination}
\label{sec:parallel}

The elementary fact that, if $\bfV\bfJ_i\bfV^\top<\zeros$ for a
set of subsystem Jacobian matrices $\bfJ_i$, then $\bfV\left(\sum_i
  \alpha_i(t)\bfJ_i \right)\bfV^\top<\zeros$ for positive
$\alpha_i(t)$, can be used in many ways. Note that it does not
represent a combination of different groups as in the above
paragraphs, but rather a superposition of different dynamics within
one group.

One such interpretation, as in \cite{Slo} for contracting systems, is
to assume that for a given system $\dot{\bfx} = \bff(\bfx,t)$, several
types of additive couplings $\bfL_i(\bfx,t)$ lead stably to the {\it
  same} invariant set, but to different synchronized behaviors. Then
any convex combination ($\alpha_i(t) \ge 0,\sum_i \alpha_i(t) = 1$) of
the couplings will lead stably to the same invariant set. Indeed,
$$
\bff(\bfx,t) - \sum_i \alpha_i(t)\bfL_i(\bfx,t)= \sum_i \alpha_i(t)
\left[\bff(\bfx,t) - \bfL_i(\bfx,t)\right]< \zeros
$$

The $\bfL_i(\bfx,t)$ can be viewed as {\it synchronization primitives}
to shape the behavior of the combination.


\section{Examples}

In conclusion, let us briefly discuss some general directions of
application of the above results to a few classical problems in
systems neuroscience and robotics.

\subsection{Coincidence detection for multiple groups} 

\label{sec:manies}

Coincidence detection is a classic mechanism proposed for segmentation
and classification.  In an image for instance, elements moving at a
common velocity are typically interpreted as being part of a single
object, and this even when the image is only composed of random
dots~\cite{Llinas,Koch}.

As mentioned in section \ref{sec:typology}, the possibility of
decentralized synchronization via central diffusive couplings can be
used in building a coincidence detector. In~\cite{WangSlo5}, inspired
in part by~\cite{Brody}, the authors consider a leader-followers
network of FitzHugh-Nagumo\footnote{See appendix
  \ref{sec:oscillators}.} oscillators, where each follower oscillator
$i$ (an element of $S_2$, see section \ref{sec:diffusive}) receives an
external input $I_i$ as well as a diffusive coupling from the leader
oscillator (the element $e_1$ of $S_1$). Oscillators $i$ and $j$
receiving the same input ($I_i=I_j$) synchronize, so that choosing the
system output as $\sum_{1\leq i \leq n} [\dot{v}_i]^+$ captures the
moment when a large number of oscillators receive the same input.

However, the previous development also implies that this very network
can detect the moments when \emph{several} groups of identical inputs
exist. Furthermore, it is possible to identify the number of such
groups and their relative size. Indeed, assume that the inputs are
divided into $k$ groups, such that for each group $S_m$, one has
$\forall i,j\in S_m, I_i=I_j$. Since the oscillators in $S_m$ only
receive as input (a) the output of the leader, which is the same for
everybody and (b) the external input $I_i$, which is the same for
every oscillator in group $S_m$, they are input-symmetric and should
synchronize with each other (cf. section \ref{sec:syminv} and section
\ref{sec:diffusive}).

\begin{figure}[ht]
  \begin{minipage}[ht]{1\textwidth}
    \centering
    \includegraphics[scale=0.3]{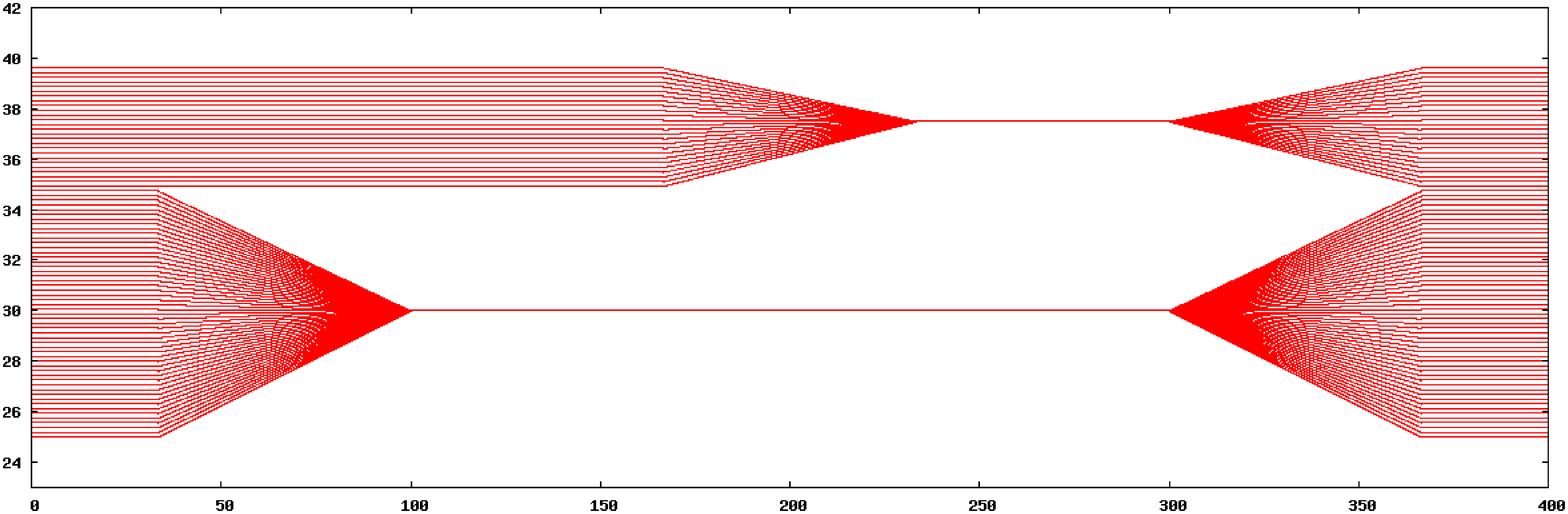}
  \end{minipage}\\[5pt]
  \begin{minipage}[ht]{1\textwidth}
    \centering
    \includegraphics[scale=0.3]{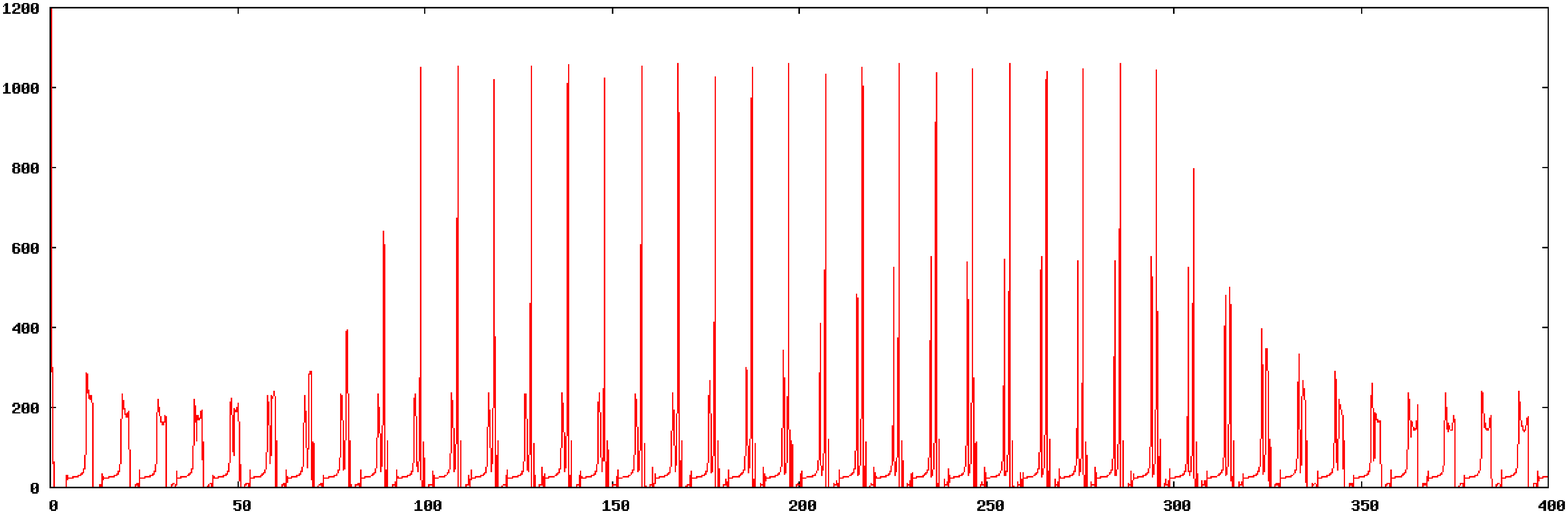}
  \end{minipage}
  \caption[]{\small Simulation results for the coincidence
    detector. The network is composed of one leader with a constant
    input $I_0=25$, and 80 followers whose inputs vary with time as
    shown in the upper figure. The lower figure plots the system output
    $\left(\sum_{1\leq i \leq 80} [\dot{v}_i]^+\right)$ against time. 

    One can clearly observe the existence of two successive, well
    separated spikes per period in a time interval around $t=250$.
    Furthermore, one spike is about twice as large as the other one.
    This agrees with the inputs, since around $t=250$, they are
    divided into two groups : 1/3 of them with value 37, 2/3 with
    value 30.  }
  \label{fig:mae_simu}
\end{figure}

Some simulation results are shown in figure \ref{fig:mae_simu}. Note
that contrast between groups could be further enhanced by using
nonlinear ``synapses'', e.g. introducing input-dependent delays, which
would preserve the symmetries. Similarly, any feedback mechanism to
the leader oscillator would also preserve the input-symmetries.

Finally, adding all-to-all identical connections between the follower
oscillators would preserve the input-symmetries and further increase
the convergence rate, but at the price of vastly increased complexity.

\subsection{Fast symmetry detection}

Symmetry, in particular bilateral symmetry, has also been shown to
play a key role in human perception \cite{Braitenberg}. Consider a
group of oscillators having the same individual dynamics and connected
together in a symmetric manner. If we present to the network an input
having the same symmetry, some of the oscillators will synchronize as
predicted by the theoretical results of section \ref{sec:syminv}.

\begin{figure}[ht]
  \begin{minipage}[h]{0.29\textwidth}
    \centering    
    \includegraphics[scale=0.35]{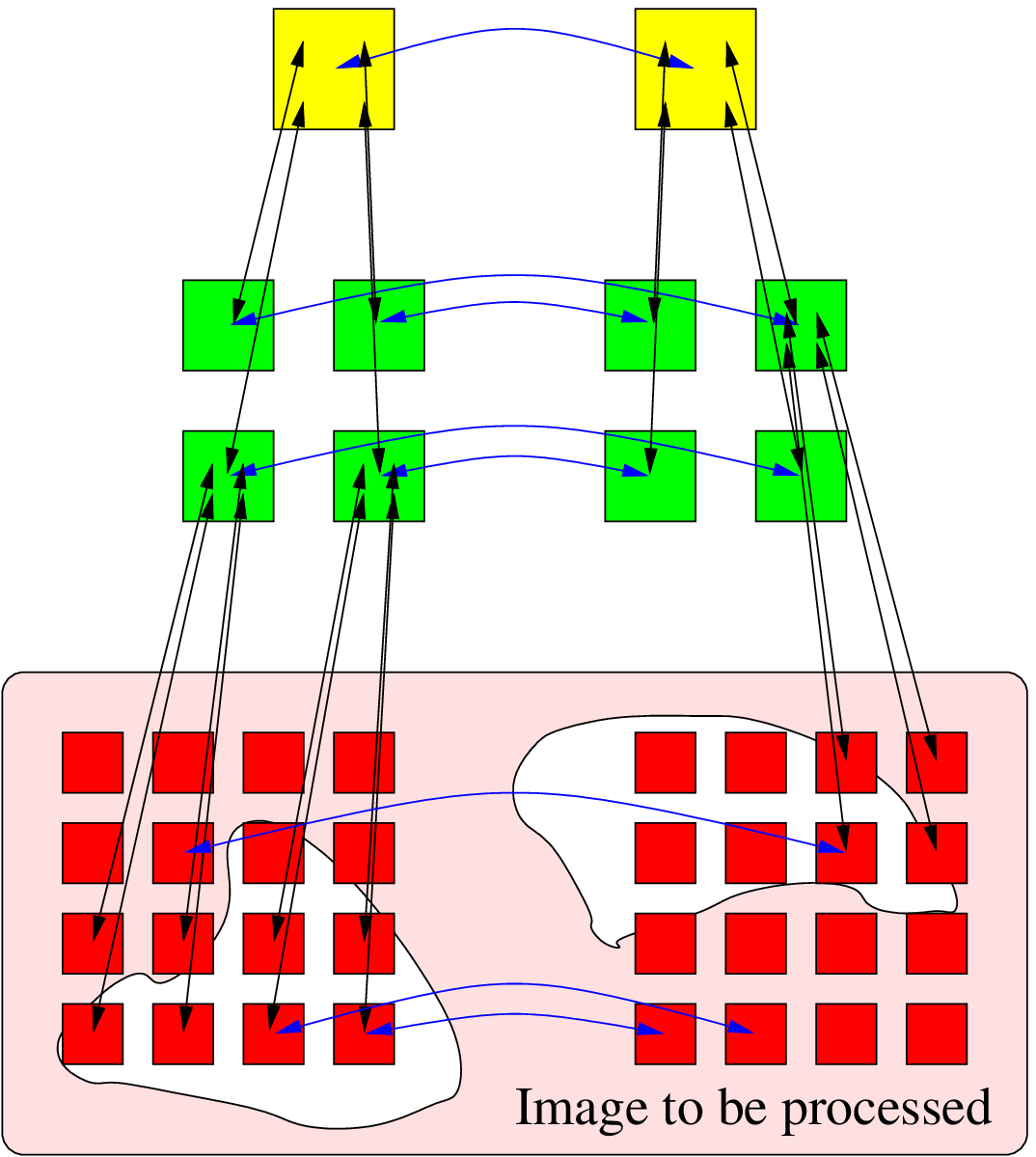}
  \end{minipage}
  \begin{minipage}[h]{0.7\textwidth}
    \flushleft
    \caption[]{\small A fast bilateral symmetry detector. \\
      Every oscillator has the
      same dynamics as its mirror image and the two are coupled through
      diffusive connection. The inputs are provided to the detector
      through the bottom (red) layer. Then ``information'' travels
      bottom-up : each layer is connected to the layer right above
      it. Top-down feedback is also possible. \\
      Assume now that a mirror symmetric image is submitted to the
      network. The network, which is mirror symmetric by construction,
      now receives a mirror symmetric input. Thus, the concurrent
      synchronization subspace where each oscillator is exactly in the
      same state as its mirror image oscillator is
      flow-invariant. Furthermore, the diffusive connections, if they
      are strong enough (see 2.2), guarantee contraction on the
      orthogonal space. By using the theoretical results above, one
      can deduce the exponential convergence to the concurrent
      synchronization regime. In particular, the difference between
      the top two oscillators should converge exponentially to zero.} 
    \label{fig:detector}
  \end{minipage}
\end{figure}

\begin{figure}[ht]
  \begin{minipage}[h]{0.29\textwidth}
    \centering    
    \includegraphics[scale=0.6]{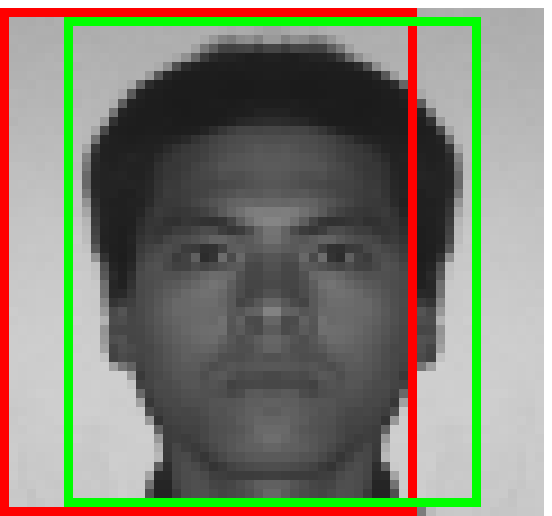}
  \end{minipage}
  \begin{minipage}[h]{0.7\textwidth}
    \flushleft
    \caption[]{\small Simulation on an artificial image. \\
      We create a $56\times 60$ pixels symmetric image from a real
      picture of one of the authors. We give it as input to a network
      similar to the one in figure~\ref{fig:detector}. The first
      (bottom) layer of the network is composed of $7\times 6= 42$
      FitzHugh-Nagumo oscillators (21 pairs)  
      each receiving the sum of the intensities of $8\times 8= 64$
      pixels, thus covering at every instant an active window of
      $56\times 48$ pixels. The second layer consists of 4
      oscillators, each receiving inputs from 9 or 12 oscillators of 
      the first layer. The third layer is composed of 2 oscillators.  
      
      At $t=0$, the active window is placed on the left of the image
      (red box) and, as $t$ increases, it slides towards the right. At
      $t=T/2$, where $T$ is the total time of the simulation, the
      position of the window is exactly at the center of the image
      (green box) (see the simulation results in figure
      \ref{fig:symres}).}
    \label{fig:symsimu}
  \end{minipage}
\end{figure}

\begin{figure}[ht]
  \centering
  \includegraphics[scale=0.35]{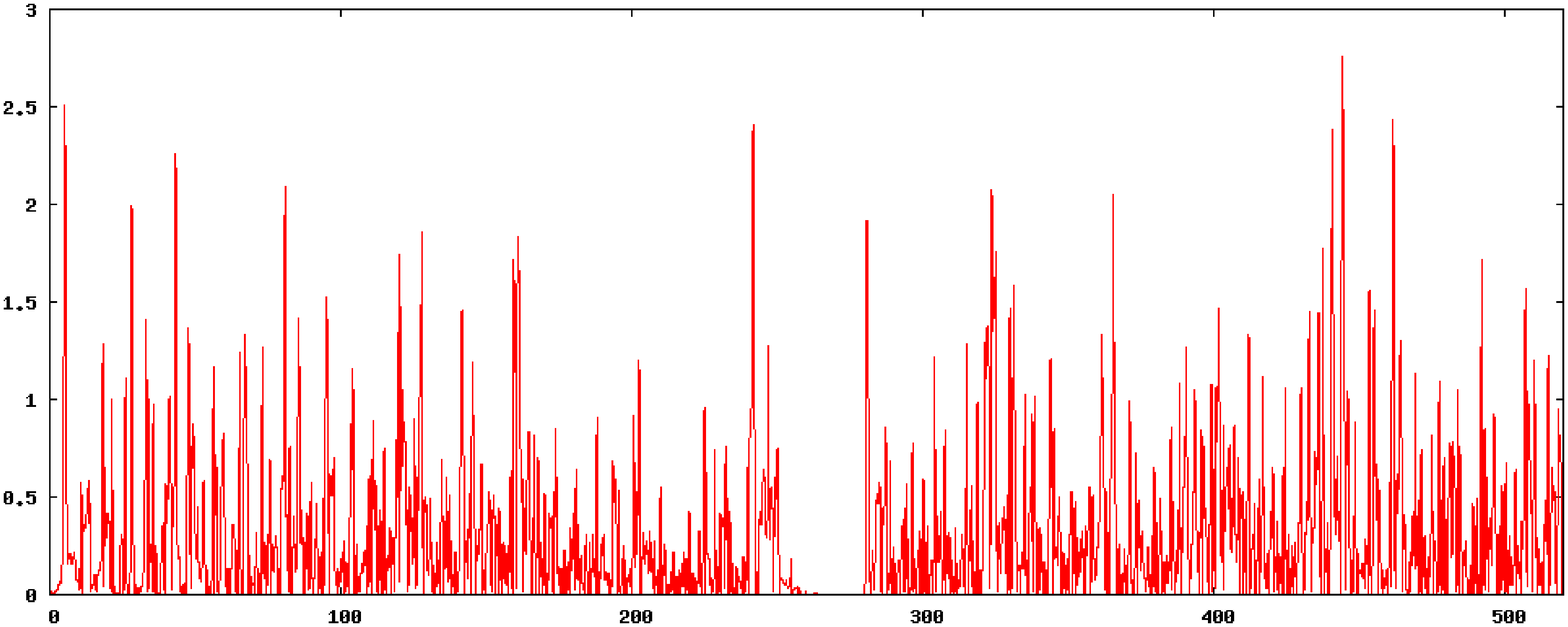}
  \caption[]{\small Result for the simulation of figure 
    \ref{fig:symsimu}. The figure shows $|v_1-v_2|$ 
    where $v_1$ and $v_2$ are the voltages of the two FN oscillators
    in the top layer. One can clearly observe that, around $t=T/2$
    ($T=520$ in this simulation), there is a short time interval
    during which the two oscillators are fully synchronized.}
  \label{fig:symres}
  \end{figure}

One application of this idea is to build a fast bilateral symmetry
detector (figures \ref{fig:detector}, \ref{fig:symsimu},
\ref{fig:symres}), extending the oscillator-based coincidence
detectors of the previous section. Although based on a radically
different mechanism, this symmetry detector is also somewhat
reminiscent of the device in~\cite{Braitenberg}.

Some variations are possible :

\begin{enumerate}
\item \emph{Other types of invariance.}  It is easy to modify the
  network in order to deal with multi-order (as opposed to bilateral)
  symmetry, or other types of invariance (translation, rotation,
  \dots). In each case, the network should have the same invariance
  pattern as what it is supposed to detect.
\item Since the exponential convergence rate is known, the network may
  be used to track \emph{time-varying} inputs, as in the coincidence
  detection algorithm of \cite{WangSlo5}.
\item \emph{Multidimensional inputs.}  Coincidence detectors and
  symmetry detectors may also handle multidimensional inputs. Two
  approaches are possible. One can either ``hash'' each
  multidimensional input into a one-dimensional input, and give the
  set of so-obtained one-dimensional inputs to the network. Or one can
  process each dimension independently in separate networks and then
  combine the results in a second step.
\end{enumerate}

\subsection{Central pattern generators} 

\label{sec:cpg}

In an animal/robotics locomotion context, central pattern generators
are often modelled as coupled nonlinear oscillators delivering
phase-locked signals. We consider here a system of three coupled
2-dimensional Andronov-Hopf oscillators \cite{IzhiKura}, very similar
to the ones used in the simulation of salamander locomotion \cite{Ijs}
:
\[
\left\{
\begin{array}{l}
\dot{\bfx}_1=\bff(\bfx_1)+k(\bfR_\frac{2\pi}{3}\bfx_2-\bfx_1)\\
\dot{\bfx}_2=\bff(\bfx_2)+k(\bfR_\frac{2\pi}{3}\bfx_3-\bfx_2)\\
\dot{\bfx}_3=\bff(\bfx_3)+k(\bfR_\frac{2\pi}{3}\bfx_1-\bfx_3)
\end{array}\right.
\]
where $\bff$ is the dynamics of an Andronov-Hopf oscillator and the
matrix $\bfR_\frac{2\pi}{3}$ describes a $\frac{2\pi}{3}$ planar
rotation :
\[
\bff\left(\begin{array}{l}x\\y\end{array}\right)=
\left(\begin{array}{l}
    x-y-x^3-xy^2\\
    x+y-y^3-yx^2
\end{array}\right) \quad \quad  \quad \quad \quad
\bfR_\frac{2\pi}{3}= \left(\begin{array}{ll}
    -\frac{1}{2} & -\frac{\sqrt{3}}{2}\\
    \frac{\sqrt{3}}{2} & -\frac{1}{2}\\
  \end{array}\right)
\]

We can rewrite the dynamics as
$\dot{\bfxg}=\bffg(\bfxg)-k\bfL\bfxg$, where
\[
\bfL=\left(\begin{array}{rrr}
    \bfI_2&-\bfR_\frac{2\pi}{3}&\zeros \\
    \zeros& \bfI_2&-\bfR_\frac{2\pi}{3} \\
    -\bfR_\frac{2\pi}{3}& \zeros & \bfI_2
\end{array}\right)
\]

First, observe that the \emph{linear} subspace $\sM=
\left\{\left(\bfR_\frac{2\pi}{3}^2(\bfx),\bfR_\frac{2\pi}{3}(\bfx),\bfx\right):
  \bfx\in\mathbb{R}^2\right\}$ is flow-invariant\footnote{As it is
  suggested in footnote 7, the flow-invariance of $\sM$ can be
  understood here as being ``created'' by the internal symmetries of
  the oscillators' dynamics.}, and that $\sM$ is also a subset of
$\mathrm{Null}(\bfL_s)$.  Next, remark that the characteristic
polynomial of $\bfL_s$ is $X^2\left(X-3/2\right)^4$ so that the
eigenvalues of $\bfL_s$ are 0, with multiplicity 2, and 3/2, with
multiplicity 4. Now since $\sM$ is 2-dimensional, it is exactly the
nullspace of $\bfL_s$, which implies in turn that $\sM^\perp$ is the
eigenspace corresponding to the eigenvalue 3/2.

\begin{figure}[ht]
  \begin{minipage}[ht]{0.24\textwidth}
    \centering
    \includegraphics[scale=0.37]{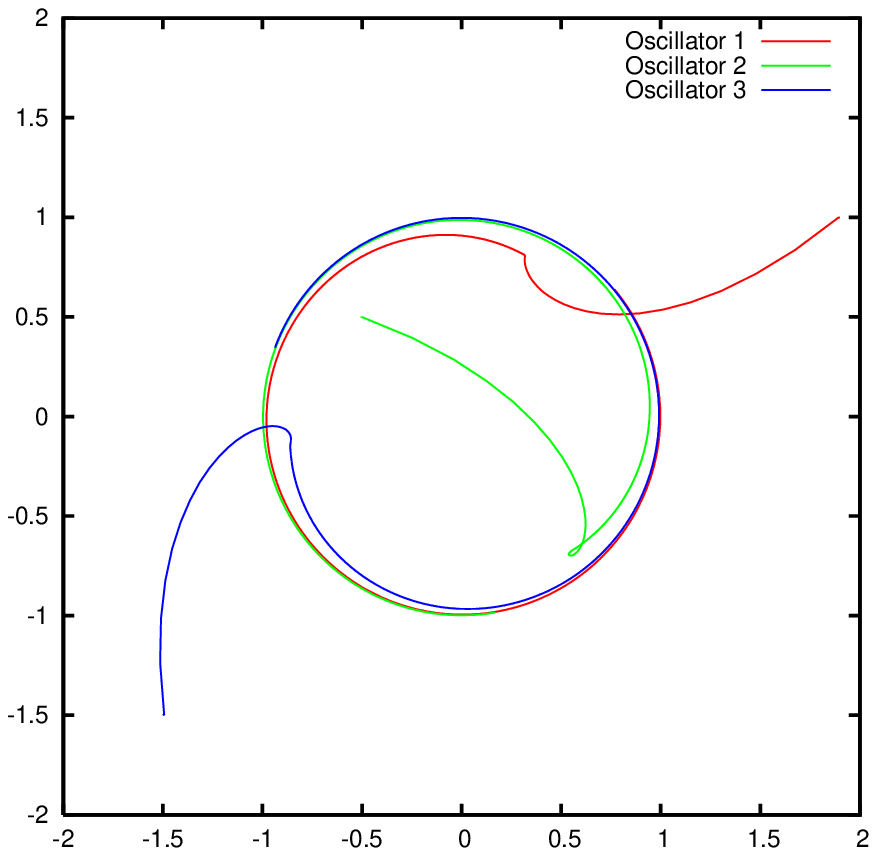}
  \end{minipage}
  \begin{minipage}[ht]{0.24\textwidth}
    \centering
    \includegraphics[scale=0.37]{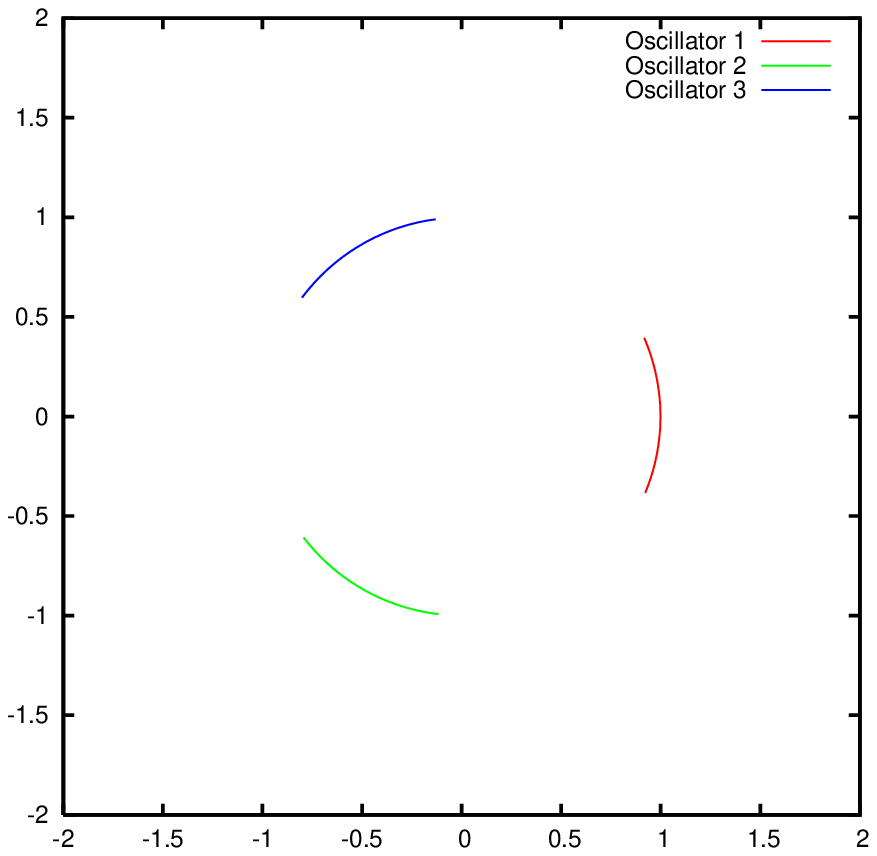}
  \end{minipage}
  \begin{minipage}[ht]{0.25\textwidth}
    \centering
    \includegraphics[scale=0.37]{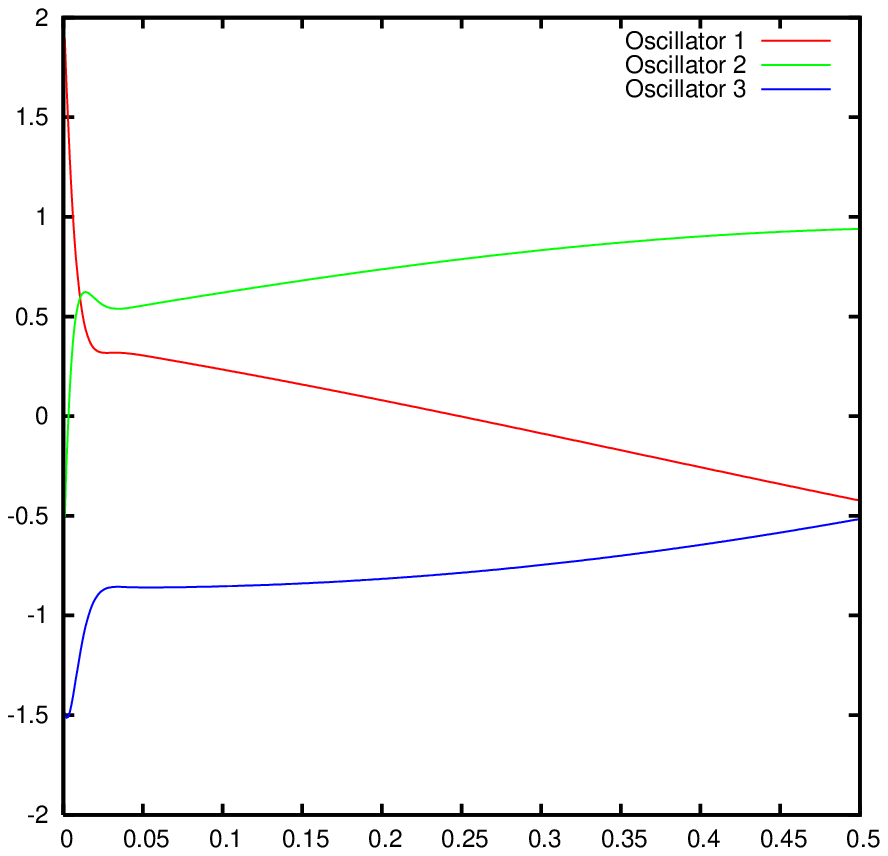}
  \end{minipage} 
   \begin{minipage}[ht]{0.25\textwidth}
    \centering
    \includegraphics[scale=0.37]{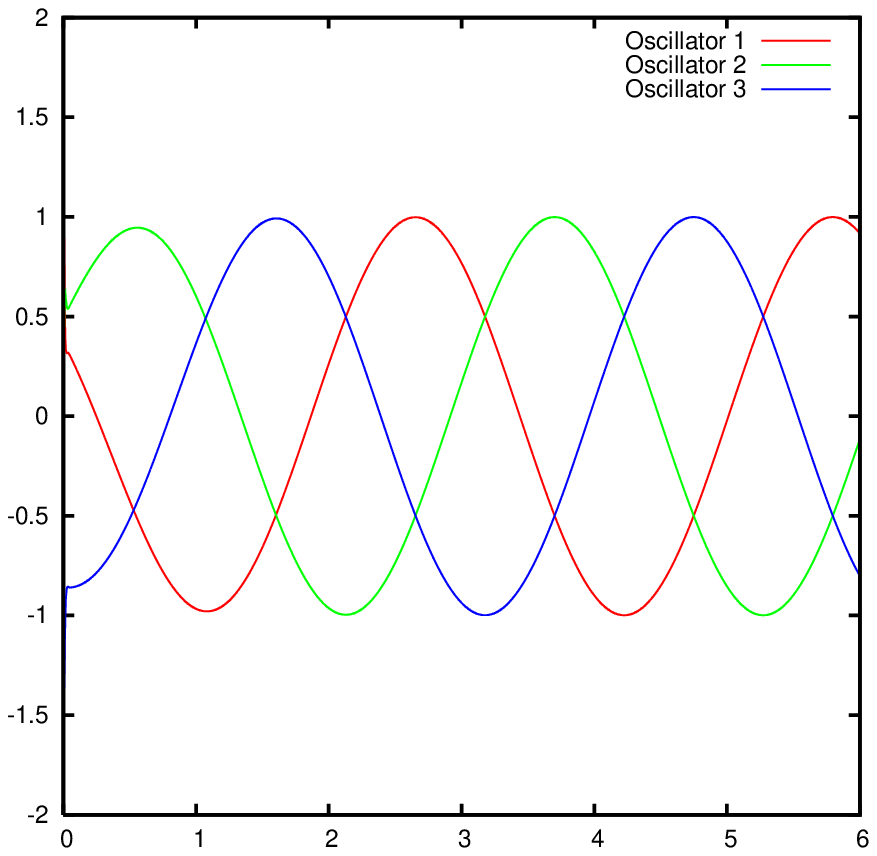}
  \end{minipage}  
  \caption[]{\small Simulation for three coupled Andronov-Hopf
    oscillators. In the first two figures, we plot $y_i$ against $x_i$
    for $1\leq i \leq 3$. Figure a shows the behavior of the
    oscillators for $0\leq t \leq 3s$, figure b for $5.6s\leq t \leq
    6s$. In figures c and d, we plot $x_1$, $x_2$, $x_3$ against time.
    Figure c for $0\leq t \leq 0.5s$, figure d for $0\leq t \leq 6s$.}
  \label{fig:ah}
\end{figure}

Moreover, the eigenvalues of the symmetric part of
$\frac{\partial{\bffg}}{\partial{\bfxg}}(x,y)$ are $1-(x^2+y^2)$ and
$1-3(x^2+y^2)$, which are upper-bounded by 1. Thus, for $k>2/3$ (see
equation (\ref{equ:synccond}) in section \ref{sec:global}), the three
systems will \emph{globally exponentially converge to a
  $\pm\frac{2\pi}{3}$-phase-locked state} (i. e. a state in which the
difference of the phases of two consecutive elements is constant and
equals $\pm\frac{2\pi}{3}$). A computer simulation is presented in
figure \ref{fig:ah}.

\textbf{Relaxing the symmetry or the diffusivity condition :} In the
previous example, the flow-invariance of the phase-locked state is due
to (a) the internal symmetry of the individual dynamics $\bff$, (b)
the global symmetry of the connections and (c) the ``diffusivity'' of
the connections (of the form $k(\bfR\bfx_2-\bfx_1)$). Observe now, as
in section \ref{sec:syminv}, that this flow-invariance can be
preserved when one out of the two conditions (b) and (c) is relaxed.
Consider for example the two following systems :
\begin{itemize}
\item Symmetric but not ``diffusive'' :
\[
\left\{\begin{array}{l}
    \dot{\bfx}_1=\bff(\bfx_1)+k\bfR_\frac{2\pi}{3}\bfx_2\\
    \dot{\bfx}_2=\bff(\bfx_2)+k\bfR_\frac{2\pi}{3}\bfx_3\\
    \dot{\bfx}_3=\bff(\bfx_3)+k\bfR_\frac{2\pi}{3}\bfx_1
\end{array}\right.
\]
(the connections are ``excitatory-only'' in the sense of
section~\ref{sec:excit}).
\item ``Diffusive'' but not symmetric :
\[
\left\{\begin{array}{l}
    \dot{\bfx}_1=\bff(\bfx_1)+k_1(\bfR_1\bfx_2-\bfx_1)\\
    \dot{\bfx}_2=\bff(\bfx_2)+k_2(\bfR_2\bfx_3-\bfx_2)\\
    \dot{\bfx}_3=\bff(\bfx_3)+k_3(\bfR_3\bfx_1-\bfx_3)
  \end{array}\right.
\]
where the $\bfR_i$ represent any planar rotations such that
$\bfR_1\bfR_2\bfR_3 = \bfI_2$ (i.e., any arbitrary phase-locking).
\end{itemize}

By keeping in mind that for any planar rotation $\bfR$ and state
$\bfx$, one has $\bff(\bfR\bfx)=\bfR(\bff(\bfx))$, it is immediate to
show the flow-invariance of
$\left\{\left(\bfR_\frac{2\pi}{3}^2(\bfx),\bfR_\frac{2\pi}{3}(\bfx),\bfx\right):
  \bfx\in\mathbb{R}^2\right\}$ in the first case, and of
$\left\{\left(\bfR_1\bfR_2(\bfx),\bfR_1(\bfx),\bfx\right):
  \bfx\in\mathbb{R}^2\right\}$ in the second case. Note however that
the computations of the projected Jacobian matrices are different, and
that in the first case the limit cycle's radius varies with $k$ (cf.
section \ref{sec:excit}).

Finally, note that

\begin{enumerate}
\item All the results of this section can be immediately extended to
  systems with more oscillators.
\item As compared to results based only on phase oscillators, this
  analysis guarantees global exponential convergence, rather than
  assuming that synchronization is already essentially achieved. In
  addition, it exhibits none of the topological difficulties that may
  arise when coupling large numbers of phase oscillators.
\item If $\bff$ is less symmetric, only connections that exhibit
  the same symmetry as $\bff$ can lead to a non-trivial
  flow-invariance subspace. 
\item It is also possible to extend this study to systems composed of
  oscillators with larger dimensions (living in $\mathbb{R}^3$ for
  example), although a locomotion interpretation may be less relevant.
\end{enumerate}

\subsection{Filtered connections and automatic gait selection}

Replacing ordinary connections in the CPG described in section~\ref{sec:cpg}
by filters enables \emph{frequency-based symmetry selection}. This
idea may have powerful applications, one of which could be automatic
gait selection in locomotion.

\begin{figure}[ht]
  \begin{minipage}[h]{0.4\textwidth}
    \centering    
    \includegraphics[scale=0.6]{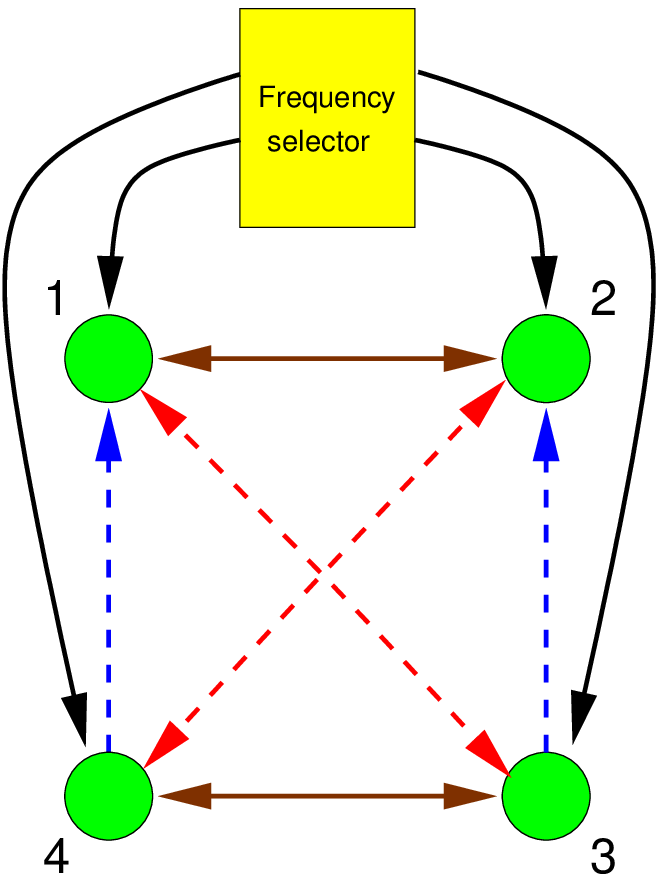}
  \end{minipage}
  \begin{minipage}[h]{0.59\textwidth}
    \flushleft
    \caption[]{\small A CPG with filtered connections. \\
      The connections from the command box set the same frequency for
      the four oscillators. The $1\leftrightarrow 2$ and
      $3\leftrightarrow 4$ arrows represent permanent
      \emph{anti-synchronization} connections (i.e. connection $j\to
      i$ is of the form $k(-\bfx_j-\bfx_i)$). The $1\leftrightarrow 3$
      and $2\leftrightarrow 4$ arrows represent \emph{synchronization}
      connections and they are \emph{high-pass} filtered. Finally, the
      $3\to 2$ and $4\to 1$ arrows stand for \emph{quarter-period
        delay} connections (i.e. connection $j\to i$ is of the form
      $k(\bfR_{-\frac{\pi}{2}}\bfx_j-\bfx_i)$, see section
      \ref{sec:cpg}) and they are \emph{low-pass} filtered.}
    \label{fig:cpg}
  \end{minipage}
\end{figure}

Consider for example the mechanism described in figure \ref{fig:cpg}.
At low frequencies, the $1\leftrightarrow 3$ and $2\leftrightarrow 4$
connections are filtered out, so that the actual connections are
$1\leftrightarrow 2$ and $3\leftrightarrow 4$ (anti-synchronization)
and $3\to 2$ and $4\to 1$ (quarter-period delay). The only non-trivial
flow-invariant subspace is then
$\ \{\bfx_1=\bfR_\frac{\pi}{2}(\bfx_3)=-\bfx_2=\bfR_\frac{3\pi}{2}(\bfx_4)\}
$.
On the contrary, the $3\to 2$ and $4\to 1$ connections are filtered
out at high frequencies, so that the flow-invariant subspace becomes
$\ \{\bfx_1=\bfx_3=-\bfx_2=-\bfx_4\}$.

Similarly to section \ref{sec:cpg}, strong enough coupling gains
ensure convergence to either of these two subspaces, according to the
frequency at which the oscillators are running. Note that standard
techniques allow sharp causal filters with frequency-independent
delays to be constructed easily~\cite{Oppenheim}.

An analogy with horse gaits could be made in this simplified setup, by
associating the low-frequency regime with the \emph{walk} (left fore,
right hind, right fore, left hind), and the high-frequency regime with
the \emph{trot} (left fore and right hind simultaneously, then right
fore and left hind simultaneously). Transitions between the two
regimes would occur automatically according to the speed of the horse
(the frequency of its gait).

\subsection{Temporal binding}

The previous development has suggested a mechanism for stable
accumulation and interaction of concurrently synchronized groups,
showing that the simple conditions for contraction to a linear
subspace, combined with the high fan-out of typical neurons, increased
the plausibility of large concurrently synchronized structures being
created in the central nervous system in the course of evolution and
development. The recently established pervasiveness of electrical
synapses~\cite{fukuda06} would also be consistent with such
architectures.

More speculatively, different ``rhythms'' ($\alpha, \beta, \gamma,
\delta$) are known to coexist in the brain, which, in the light of the
previous analysis, may be interpreted and modelled as concurrently
synchronized regimes. Since contracting systems driven by periodic
inputs will have states of the same period~\cite{LohSlo}, different
but synchronized computations could be robustly carried out by
specialized areas in the brain using synchronized elements as their
inputs. Such a temporal
``binding''~\cite{SinGra,Grossberg,Llinas,Tononi,Koch,NieEbi,YazGro,Crick,fukuda06}
mechanism would also complement the general argument in~\cite{SloLoh}
that multisensory integration may occur through the interaction of
contracting computational systems connected through an extensive
network of feedback loops. In this context, and along the lines of
section~\ref{sec:typology}, a translation to concurrent
synchronization of recent results on centralized contracting
combinations~\cite{TabSlo} may be particularly relevant. Making these
observations precise is the subject of future research.





\section*{Acknowledgments} 

We are grateful to Jake Bouvrie, Nicolas Tabareau and Sacha Zyto for
stimulating discussions, and to an anonymous reviewer for thoughtful
suggestions to improve the presentation. The first author would like
to thank CROUS, Paris for financial support.

\appendix

\section{FitzHugh-Nagumo oscillators}   

\label{sec:oscillators}

Some of our simulations involve coupled FitzHugh-Nagumo neural
oscillators \cite{Fitz,Nagumo} 
\[
\left\{\begin{array}{l}
    \dot{v}_i=v_i(\alpha-v_i)(v_i-1)-w_i+I_i+k(v_0-v_i)\\
    \dot{w}_i=\beta v_i-\gamma w_i
\end{array}\right. \quad 1\leq i \leq n
\]
In this paper, we use the following parameters values : $\alpha=6,\
\beta=3,\ \gamma=0.09$. 

The contraction analysis of FitzHugh-Nagumo oscillators can be adapted
from \cite{WangSlo}.










\begin{thebibliography}{00}

\bibitem{Brody} C. Brody, J. Hopfield. Simple Networks for
  Spike-Timing-Based Computation, with Application to Olfactor
  Processing. \emph{Neuron}, {\bf{37}} (5):843--852, 2003.

\bibitem{Belykh1} V. Belykh, I. Belykh, M. Hasler, K. Nevidin. Cluster
  Synchronization in Three-dimensional Lattices of Diffusively Coupled
  Oscillators. \emph{Int. J. of Bifurcation and Chaos},
  {\bf{13}}:755--779, 2003.

\bibitem{Braitenberg} V. Braitenberg. \emph{Vehicles: Experiments in
    Synthetic Psychology}, chap. 9. The MIT Press, 1984.

\bibitem{Crick} F. Crick, C. Koch. What is the Function of the
  Claustrum. \emph{Phil. Trans. Roy. Soc. Lond. B},
  {\bf{360}}:1271--1279, 2005.

\bibitem{DayHin} P. Dayan, G. Hinton, R. Neal, R. Zemel.  The
  Helmholtz Machine. \emph{Neural Computation}, {\bf{7}}, 1995.

\bibitem{DioGolSte} B. Dionne, M. Golubitsky, I. Stewart. Coupled
  cells with internal symmetry. \emph{Nonlinearity},
  \textbf{9}:559--599, 1996.

\bibitem{Fiedler} M. Fiedler. Algebraic Connectivity of Graphs.
  \emph{Czechoslovak Mathematical Journal}, 1976.

\bibitem{Fitz} R. FitzHugh. Impulses and Physiological States in
  Theoretical Models of Nerve Membrane. \emph{Biophys. Journal},
  {\bf{1}}:445-466, 1961.


\bibitem{fukuda06} T. Fukuda, T. Kosaka, W. Singer and R.A.W. Galuske.
  Gap Junctions among Dendrites of Cortical GABAergic Neurons
  Establish a Dense and Widespread Intercolumnar Network. {\emph{The
      Journal of Neuroscience}}, {\bf{26}} (13):3434--3443, 2006.

\bibitem{GeoHaw} D. George, J. Hawkins. Invariant Pattern Recognition
  using Bayesian Inference on Hierarchical Sequences, Stanford
  University, 2005.

\bibitem{GolSte} M. Golubitsky, I. Stewart. Synchrony versus Symmetry
  in Coupled Cells. \emph{Equadiff 2003: Proceedings of the
    International Conference on Differential Equations}.

\bibitem{GolSteTor} M. Golubitsky, I. Stewart, A. T\"or\"ok. Patterns
  of Symmetry in Coupled Cell Networks with Multiple Arrows.
  \emph{SIAM J. Appl. Dynam. Sys.}, {\bf{4}} (1):78--100, 2005.


\bibitem{Grossberg} S. Grossberg. The Complementary Brain : a Unifying
  View of Brain Specialization and Modularity. \emph{Trends in
    Cognitive Sciences}, {\bf{4}}:233--246, 2000.

\bibitem{horn} R. Horn, C. Johnson. \emph{Matrix Analysis}.  Cambridge
  University Press, 1985.

\bibitem{Ijs} A. Ijspeert, A. Crespi, J.-M. Cabelguen. Simulation and
  Robotic Studies of Salamander Locomotion : Applying Neurobiological
  Principles to the Control of Locomotion in Robots, 2005.

\bibitem{Izhi} E. Izhikevich. Simple Model of Spiking Neuron.
  \emph{IEEE Trans. on Neural Networks}, {\bf{14}}(6):1569--1572,
  2003.

\bibitem{IzhiDes} E. Izhikevich, N. Desai, E. Walcott, F.
  Hoppensteadt. Bursts as a Unit of Neural Information : Selective
  Communication via Resonance.  \emph{Trends in Neuroscience},
  {\bf{26}} (3):161--167, 2003.

\bibitem{IzhiKura} E. Izhikevich, Y. Kuramoto. Weakly Coupled
  Oscillators. \emph{Encyclopedia of Mathematical Physics}. Elsevier,
  2006.

\bibitem{JadLinMo} A. Jadbabaie, J. Lin, A. Morse. Coordination of
  Groups of Mobile Autonomous Agents using Nearest Neighbor Rules.
  \emph{IEEE Transactions on Automatic Control}, {\bf{48}}:988--1001,
  2003.

\bibitem{JadMotBar} A. Jadbabaie, N. Motee, M. Barahona. On the
  Stability of the Kuramoto Model of Coupled Nonlinear Oscillators.
  \emph{Proceedings of the American Control Conf.}, Boston, MA June
  30--July 2, 2004.

\bibitem{Kandel} E. Kandel, J. Schwartz, T. Jessel.  \emph{Principles
    of Neural Science}, 5th ed., McGraw-Hill, 2006.

\bibitem{Knu} D. Knuth. \emph{The Art of Computer Programming}, 3rd
  Ed. Addison-Wesley, 1997.

\bibitem{Koch} C. Koch. \emph{The Quest for Consciousness}. Roberts
  and Company Publishers, 2004.

\bibitem{Korner} E. K\"{o}rner, M.-O. Gewaltig, U. K\"{o}rner, A.
  Richter, T. Rodemann. A model of computation in neocortical
  architecture. \emph{Neural Networks}, {\bf{12}} (7-8):989--1005,
  1999.


\bibitem{Llinas} R. Llinas, E. Leznik, F. Urbano. Temporal Binding via
  Cortical Coincidence Detection. \emph{PNAS}, {\bf{99}} (1):449--454,
  2002.

\bibitem{LinBroFra} Z. Lin, M. Broucke, B. Francis. Local Control
  Strategies for Groups of Mobile Autonomous Agents. \emph{IEEE Trans.
    on Automatic Control}, {\bf{49}}(4):622--629, 2004.

\bibitem{LohSlo} W. Lohmiller, J.-J. Slotine. On Contraction Analysis
  for Nonlinear Systems. \emph{Automatica}, {\bf{34}} (6):671--682,
  1998.

\bibitem{LohSlo2} W. Lohmiller, J.-J. Slotine. Nonlinear Process
  Control Using Contraction Theory. \emph{A.I.Ch.E.
    Journal}, {\bf{46}}(3):588-597, 2000.

\bibitem{LueWil} M. Luettgen, A. Willsky. Likelihood Calculation for a
  Class of Multiscale Stochastic Models, with Application to Texture
  Discrimination.  \emph{IEEE Transactions on Image Processing},
  {\bf{4}}(2):194--207, 1995.


\bibitem{Mount} V. Mountcastle. The Cerebral Cortex. \emph{Harvard
    University Press}, 1998.

\bibitem{Nagumo} J. Nagumo, S. Arimoto, S. Yoshizawa. An Active Pulse
  Transmission Line Simulating Nerve Axon. \emph{Proc. Inst. Radio
    Engineers}, {\bf{50}}:2061--2070, 1962.

\bibitem{NieEbi} J. Niessing, B. Ebisch, K.E. Schmidt, M. Niessing, W.
  Singer, R.A. Galuske. Hemodynamic Signals Correlate Tightly with
  Synchronized Gamma Oscillations. \emph{Science}, {\bf{309}}
  948--951, 2005.

\bibitem{OlfMur} R. Olfati-Saber, R. Murray. Consensus Problems in
  Networks of Agents With Switching Topology and Time-Delays.
  \emph{IEEE Transactions on Automatic Control},
  {\bf{49}}(9):1520-1533, 2004.

\bibitem{Oppenheim} A. Oppenheim, R. Schafer, J. Buck. 
  \emph{Discrete-Time Signal Processing, 2nd Edition}.
  Prentice-Hall,1999.

\bibitem{PhamSloReport} Q.-C. Pham, J.-J. Slotine. Attractors.
  \emph{MIT-NSL Report 0505}, 2005.

\bibitem{PogSanNij} A. Pogromsky, G. Santoboni, H. Nijmeijer. Partial
  Synchronization : from Symmetry towards Stability. \emph{Physica D},
  {\bf{172}} (1-4):65--87, 2002.

\bibitem{RaoBal} R. Rao, D. Ballard. Predictive Coding in the Visual
  Cortex. \emph{Nature Neuroscience}, {\bf{2}} (1):9--10, 1999.

\bibitem{Rao} R. Rao. Bayesian Inference and Attention in the Visual
  Cortex. \emph{Neuroreport}, {\bf{16}}(16):1843--1848, 2005.

\bibitem{Schnitzler} A. Schnitzler, J. Gross. Normal and Pathological
  Oscillatory Communication in the Brain. \emph{Nat. Rev. Neurosci.},
  {\bf{6}}, 285--296, 2005.

\bibitem{Schoffelen} J.-M. Schoffelen, R. Oostenveld, P. Fries.
  Neuronal Coherence as a Mechanism of Effective Corticospinal
  Interaction. \emph{Science}, {\bf{308}}:111--113, 2005.


\bibitem{SheRin} A. Sherman, J. Rinzel. Model for Synchronization of
  Pancreatic I3-cells by Gap Junction
  coupling. \emph{Biophys. Journal}, \textbf{59}:547--559, 1991. 


\bibitem{SinGra} W. Singer, C.  Gray. Visual Feature Integration and
  the Temporal Correlation Hypothesis. \emph{Annu. Rev. Neurosci.},
  {\bf{18}}:555--586, 1995.

\bibitem{Slo} J.-J. Slotine. Modular Stability Tools for Distributed
  Computation and Control. \emph{Int. J. Adaptive Control and Signal
    Processing}, {\bf{17}} (6):397--416, 2003.

\bibitem{SloLoh} J.-J. Slotine, W. Lohmiller. Modularity, Evolution,
  and the Binding Problem : A View from Stability Theory. \emph{Neural
    Networks}, {\bf{14}}(2):137--145, 2001.


\bibitem{Strogatz} S. Strogatz. From Kuramoto to Crawford : Exploring
  the Onset of Synchronization in Populations of Coupled Oscillators.
  \emph{Physica D}, {\bf{143}} (1-4):1--20, 2000.

\bibitem{TabSlo} N. Tabareau, J.-J. Slotine.  Notes on Contraction
  Theory. \emph{MIT-NSL Report 0503}, 2005.

\bibitem{Tononi} G. Tononi, et al. \emph{Proc. Natl. Acad. Sci.  USA}
  {\bf{95}}:3198--3203, 1998.

\bibitem{WangSlo} W. Wang, J.-J. Slotine. On Partial Contraction
  Analysis for Coupled Nonlinear Oscillators. \emph{Biological
    Cybernetics}, {\bf{92}} (1):38--53, 2005.

\bibitem{WangSlo2} W. Wang, J.-J. Slotine. Contraction Analysis of
  Time-Delayed Communications Using Simplified Wave Variables.
  PS/0512070, 25 Dec 2005.

\bibitem{WangSlo5} W. Wang, J.-J. Slotine. Fast Computation with
  Neural Oscillators. \emph{Neurocomputing}, {\bf{63}}, 2005.

\bibitem{YazGro} A. Yazdanbakhsh, S. Grossberg. Fast Synchronization
  of Perceptual Grouping in Laminar Visual Cortical Circuits.
  \emph{Neural Networks}, {\bf{17}} (5-6):707--718, 2004.

\bibitem{Zhang} Y. Zhang, et al. Partial Synchronization and
  Spontaneous Spatial Ordering in Coupled Chaotic Systems. \emph{Phys.
    Rev. E}, {\bf{63}}:026211, 2001.


\end{thebibliography}
\end{document}